\documentclass[pre,twocolumn,showkeys,showpacs]{revtex4-1}
\usepackage{graphicx,latexsym,amssymb,amsmath,amsbsy,color}

\begin{document}
\title{Statistics of Conserved Quantities in Mechanically Stable Packings of Frictionless Disks Above Jamming}
\author         {Yegang Wu and S. Teitel}
\affiliation    {Department of Physics and Astronomy, University of Rochester, Rochester, New York 14627, USA}
\date{\today}

\begin{abstract}
We numerically simulate mechanically stable packings of soft-core, frictionless, bidisperse disks in two dimensions, above the jamming packing fraction $\phi_J$.  For configurations with a fixed isotropic global stress tensor, we compute the averages, variances, and correlations of conserved quantities (stress $\Gamma_{\cal C}$, force-tile area $A_{\cal C}$, Voronoi volume $V_{\cal C}$, number of particles $N_{\cal C}$, and number of small particles $N_{s{\cal C}}$) on compact subclusters of particles ${\cal C}$, as a function of the cluster size and the global system stress.  We find several significant differences depending on whether the cluster ${\cal C}$ is defined by a fixed radius $R$ or a fixed number of particles $M$.  We comment on the implications of our findings for maximum entropy models of jammed packings.
\end{abstract}
\pacs{05.40-d, 45.70.-n, 46.65.+g}
\maketitle


\maketitle

\section{Introduction}

As one increases the density of deformable granular particles above a critical jamming packing fraction, $\phi_J$, the system undergoes a transition from a liquid-like to a solid-like state \cite{Liu+Nagel}.  
For large and massive particles, thermal fluctuations are irrelevant, and in the absence of any mechanical agitation, the dense system relaxes into a mechanically stable {\em rigid} but {\em disordered} configuration. Numerous works have considered how the global properties of such static packings scale as one approaches the jamming transition from above, $\phi\to\phi_J$ \cite{Liu+Nagel,OHern,Wyart,Ellenbroek,vanHecke}.  Here we consider the statistical properties of conserved quantities defined on finite sized subclusters of particles of the total system.  By ``conserved quantity" we mean an extensive observable which is additive over disjoint subclusters of the system, and for which the total system has a fixed value in the ensemble of configurations being considered.  Such conserved quantities have played an important role in making maximum entropy models \cite{Plischke} for the non-uniform distribution of various properties of the disordered packings \cite{Edwards,Henkes1,Henkes2,Tighe1,Tighe3,Tighe4,WuTeitel}.

We consider here a bidisperse system of soft frictionless disks in two dimensions (2D).  We will consider two different ensembles of circular clusters.  One in which the radius $R$ of the cluster is fixed and the number of particles in the cluster fluctuates; and the other in which the number of particles $M$ in the cluster is fixed and the radius fluctuates.  We find that there are several significant differences between these two ensembles: (i) For fixed $R$, averages defined on the cluster are simply related to the corresponding global parameter for any $R$; for fixed $M$, however, such averages only approach the naively expected value algebraically as the cluster size increases.  (ii) For fixed $R$, correlations between many variables decrease as the cluster size increases; for fixed $M$, however, we find that correlations appear to be constant as the cluster size increases.  We believe that these differences have important consequences for the development of maximum entropy models to describe the statistical behavior of such jammed packings.

We further investigate how the concentration of small particles in a cluster depends on cluster size, and find an algebraic variation.  We show that the notion of hyperuniformity \cite{Sollich, Zachary}, found for packings exactly at $\phi_J$, continues to hold in packings above $\phi_J$.

The conserved quantities we consider are the Voronoi volume $V$ (in our 2D system, ``volume" will be used to mean area), which has played an important role in Edwards' \cite{Edwards} statistical ensemble for jammed packings, the extensive stress $\Gamma$, which Henkes and co-workers \cite{Henkes1,Henkes2} have used to define the {\em stress ensemble}, the Maxwell-Cremona force-tile area $A$, which Tighe and co-workers \cite{Tighe1,Tighe3,Tighe4} have argued plays an important role in the distribution of pressure, as well as the number of particles $N$ and the number of small particles $N_s$.

\section{Model}

\subsection{Soft-Core Disks}

Our system is a bidisperse mixture of equal numbers of big and small circular, frictionless, disks with diameters $d_b$ and $d_s$ in the ratio $d_b/d_s=1.4$ \cite{OHern}.  If $v_{b,s}=\pi(d_{b,s}/2)^2$ is the volume of the big and small disks respectively, then the packing fraction of a system with $N$ disks in a total volume $V$ is 
\begin{equation}
\phi = \frac{N}{V}\frac{(v_s+v_b)}{2}.
\end{equation}
Disks $i$ and $j$ interact only when they overlap, with a soft-core repulsive harmonic interaction potential,
\begin{equation}
{\cal V}(r_{ij})=\left\{
\begin{array}{cl}
\frac{1}{2}k_e(1-r_{ij}/d_{ij})^2, & r_{ij}<d_{ij}\\
0, & r_{ij}\ge d_{ij}.
\end{array}
\right.
\end{equation}
Here $r_{ij}$ is the center-to-center distance between the particles, and $d_{ij}=(d_i+d_j)/2$ is the sum of their radii.  We will measure energy in units such that $k_e=1$, and length in units so that the small disk diameter $d_s=1$.

\begin{figure}[h]
\begin{center}
\includegraphics[width=1.8in]{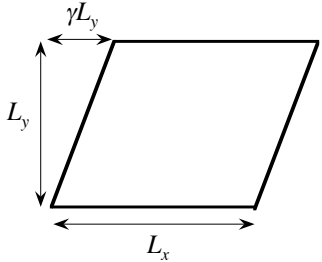}
\caption{Geometry of our system box.  $L_x$ and $L_y$ are the lengths in the $\mathbf{\hat x}$ and $\mathbf{\hat y}$ directions, and $\gamma$ is the skew ratio.  Lees-Edwards boundary conditions are used.
}
\label{f1}
\end{center}
\end{figure}

Our numerical system consists of $N=8192$ disks, which is large enough that finite size effects are negligible at the system parameters we study.  The geometry of our system box is characterized by three parameters, $L_x,L_y,\gamma$, as illustrated in Fig.~\ref{f1}; $L_x $ and $L_y$ are the lengths of the box in the $\mathbf{\hat x}$ and $\mathbf{\hat y}$ directions, while $\gamma$ is the skew ratio of the box.  We use Lees-Edwards boundary conditions \cite{LeesEdwards} to periodically repeat this box throughout all space.

\subsection{Packings With Isotropic Stress}

For this work we consider only packings with an {\em isotropic} total stress tensor,
\begin{equation}
\Sigma^{(N)}_{\alpha\beta}=\Gamma_N\delta_{\alpha\beta}, \quad \mathrm{where}\quad\Gamma_N=pV,
\end{equation}
$p$ is the system pressure, and $V=L_xL_y$ is the total system volume.   Here $\alpha,\beta$ denote the spatial coordinate directions $x,y$.  

To construct such isotropic packings, in which the shear stress vanishes, we use a scheme in which we vary the box parameters $L_x,L_y$ and $\gamma$ as we search for mechanically stable states \cite{Dagois}.
We introduce a modified energy function $\tilde U$ that depends on the particle positions $\mathbf{r}_i$, as well as $L_x,L_y,\gamma$,
\begin{equation}
\tilde U\equiv U+\Gamma_x^0\ln L_x +\Gamma_y^0\ln L_y,\quad U=\sum_{i<j}{\cal V}(r_{ij}).
\end{equation}
Here $\Gamma_x^0$ and $\Gamma_y^0$ are fixed constants representing the diagonal components of the desired diagonal stress tensor.  Noting that the interaction energy $U$ depends implicitly on the box parameters $L_x,L_y,\gamma$ via the boundary conditions, we get the relations,
\begin{equation}
\begin{aligned}
L_x\frac{\partial U}{\partial L_x}=-\Sigma_{xx}+\gamma\Sigma_{xy},&\quad\frac{\partial U}{\partial \gamma}=-\Sigma_{xy},\\ 
L_y\frac{\partial U}{\partial L_y}=-\Sigma_{yy}-\gamma\Sigma_{xy}.&
\end{aligned}
\end{equation}

Starting with randomly positioned particles within a square box of length $L$ determined by the packing fraction $\phi\approx\phi_J$, we then minimize $\tilde U$
with respect to both particle positions and box parameters.  The resulting local minimum of $\tilde U$
gives a mechanically stable configuration with force balance on each particle and a stress tensor that satisfies 
\begin{equation}
\Sigma_{xx}=\Gamma_x^0, \quad \Sigma_{yy}=\Gamma_y^0, \quad \Sigma_{xy}=0.  
\end{equation}
For isotropic states we choose $\Gamma_x^0=\Gamma_y^0=\Gamma_N$.  
For minimization we use the Polak-Ribiere conjugate gradient algorithm \cite{NR}.  We consider the minimization converged when we satisfy the condition $(\tilde U_i - \tilde U_{i+50})/\tilde U_{i+50} < 10^{-10}$, where $\tilde U_i$ is the value at the $i$th step of the minimization.  Our results are averaged over 10000 independently generated isotropic configurations.

\begin{figure}[t]
\begin{center}
\includegraphics[width=2.8in]{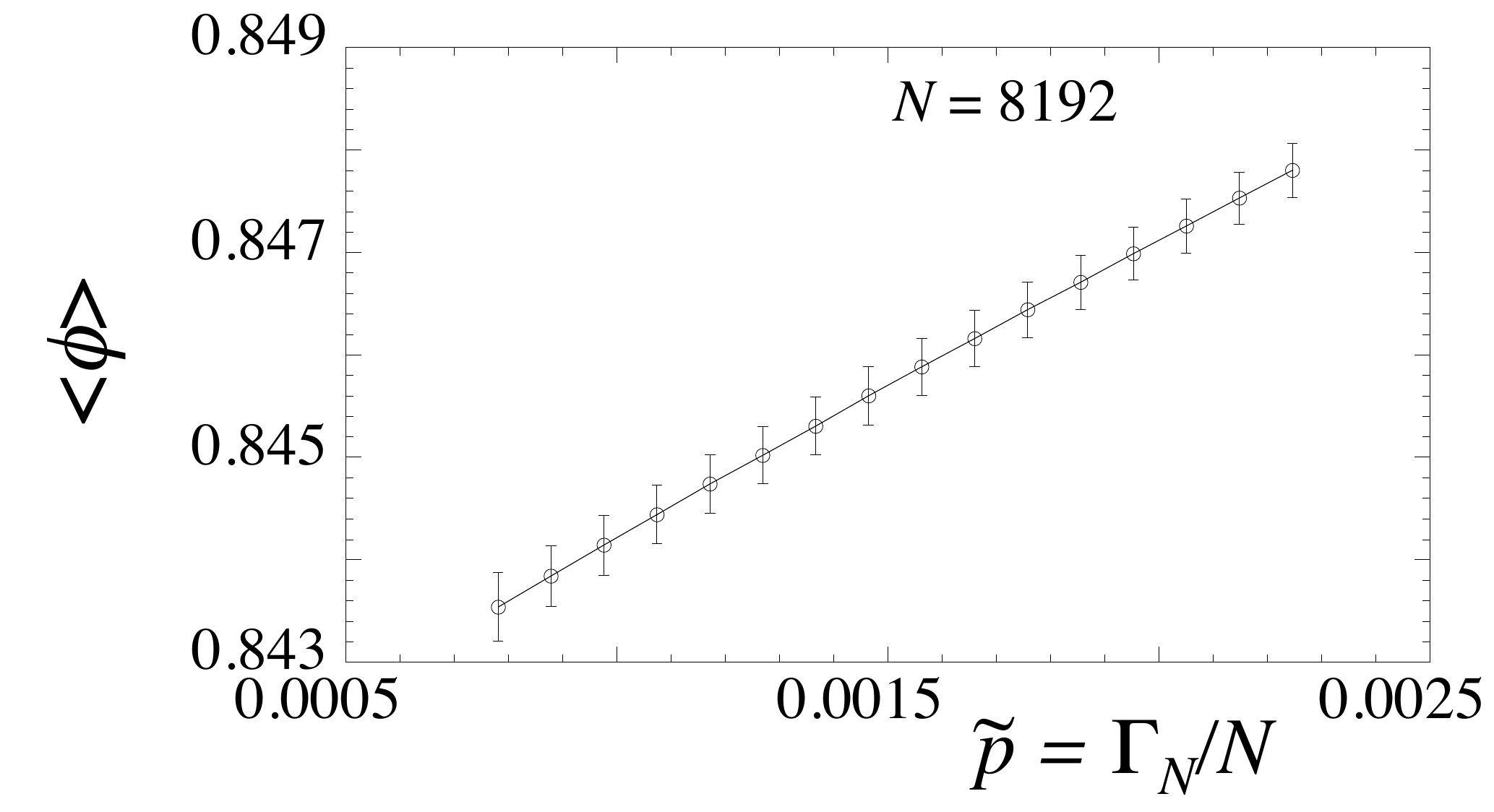}
\caption{Average packing fraction $\langle \phi\rangle$ vs total system stress per particle $\tilde p\equiv \Gamma_N/N$.  Error bars represent the width of the distribution rather than the statistical error in the average.
}
\label{f2}
\end{center}
\end{figure}

In this work we consider the range of $\Gamma_N=6.4$ to 18.4 in increments of 0.8.
Since our simulations fix both $N$ and $\Gamma_N$, we will parameterize our results by the intensive, pressure-like, variable, $\tilde p \equiv \Gamma_N/N = p(V/N)$, the total system stress per particle.  
Since our method varies the system volume $L_xL_y$ so as to achieve the desired total stress $\Gamma_N$, the packing fraction $\phi$ for fixed $\Gamma_N$ varies slightly from configuration to configuration. 
In Fig.~\ref{f2} we plot the resulting average $\langle \phi\rangle$ as a function of $\tilde p$. Error bars represent the width of the distribution of $\phi$; the relative width is roughly $0.03-0.04\%$.  The values of $\tilde p$ we consider here are all close above the jamming transition, which for our rapid quench protocol  is $\phi_J\approx 0.842$ for an infinite system \cite{VagbergFSS}.

\subsection{Cluster Ensembles}

{\em Fixed radius}: To define our clusters of particles with fixed radius $R$, we pick a point in the system at random and draw a circle of radius $R$ about that point.  All particles whose centers lie within this circle are considered part of the cluster.  For a fixed $R$, the number of particles in the cluster fluctuates.  For the clusters of radii $R=2.8$ to 8.2 considered here, the average number of particles in the cluster ranges roughly from $\langle N_R\rangle \approx 18$ to 150.  

{\em Fixed number of particles}: To define our clusters with fixed number of particles $M$, we again pick a random position in the system, draw a circle about that point, and then continuously increase the radius of the circle until we have exactly $M$ particles whose centers lie within the circle.  For such clusters, the number of particles is fixed, but the volume fluctuates.  We consider here clusters with $M=18$ to 153.

In both cases the diameters of our clusters are small enough compared to the length of the total system, that any effects of the finite total system size are negligible.

\subsection{Conserved Quantities}

{\em Stress}:  The stress tensor for a finite cluster of particles ${\cal C}$ is given by  \cite{Henkes1},
\begin{equation}
\Sigma_{\alpha\beta}^{({\cal C})}=\sum_{i\in {\cal C}}{\sum_j}^\prime {s}_{ij\alpha}{F}_{ij\beta},\quad \mathbf{F}_{ij}=-{\partial {\cal V}(r_{ij})}/{\partial\mathbf{r}_{j}}.
\end{equation}
Here $\mathbf{s}_{ij}$ is the displacement from the center of particle $i$ to its point of contact with $j$, and $\mathbf{F}_{ij}$ is the force on $j$ due to contact with $i$.
The first sum is over all particles $i$ in the cluster ${\cal C}$.  The second, primed, sum is over all particles $j$ in contact with particle $i$. The sum over all particles $i$ in the total system just gives the total stress tensor $\Sigma_{\alpha\beta}^{(N)}=\Gamma_N\delta_{\alpha\beta}$.  Since $\Sigma_{\alpha\beta}^{({\cal C})}$ is clearly additive over disjoint clusters, and its total for the entire system is constrained by $\Gamma_N$, the stress tensor is a conserved quantity.  

Although the total system stress is isotropic, the stress on any particular cluster $\Sigma_{\alpha\beta}^{({\cal C})}$ in general is not.  However the stress averaged over many independent clusters will be isotropic.  If we define
\begin{equation}
\Gamma_{\cal C}\equiv \frac{1}{2}\mathrm{Tr}[\Sigma_{\alpha\beta}^{({\cal C})}],\quad\mathrm{then}\quad
\langle\Sigma_{\alpha\beta}^{({\cal C})}\rangle=\langle\Gamma_{\cal C}\rangle\delta_{\alpha\beta}.
\end{equation}

{\em Force-tile area}:  For particles in a two dimensional mechanically stable packing, the Maxwell-Cremona force-tile for particle $i$ is obtained by rotating all its contact forces $90^\circ$ and lying them tip to tail.  Force balance then requires these to form a closed loop \cite{Ball}.  The area $A_i$ of this loop is the force-tile area.  It can be shown that such force-tiles tile space with no gaps or overlaps \cite{Tighe3}.  The tile area of a cluster of particles ${\cal C}$ is then just the sum over tile areas for each member particle, $A_{\cal C}=\sum_{i\in{\cal C}}A_i$.  The sum over all particles gives the total force-tile area $A_N$ for the entire system.  The force-tile area is thus also a conserved quantity.

\begin{figure}[h]
\begin{center}
\includegraphics[width=3in]{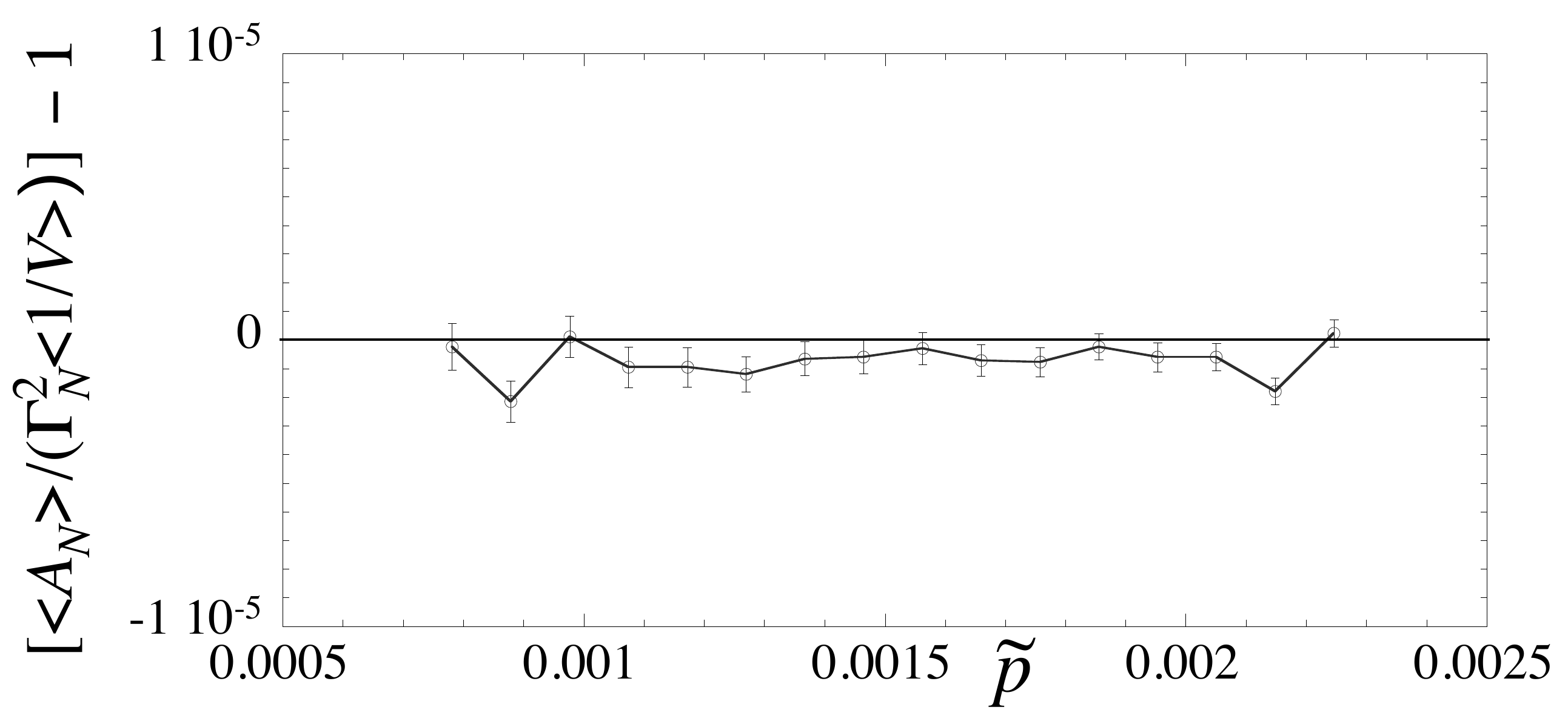}
\caption{$[\langle A_N\rangle /(\Gamma_N^2\left\langle\frac{1}{V}\right\rangle)]-1$ vs $\tilde p\equiv \Gamma_N/N$, where $A_N$ is the total force-tile area, $\Gamma_N$ the total system stress, and $V$ the total system volume.  We expect this quantity to vanish at all $\tilde p$. Our system has a total of $N=8192$ particles.
}
\label{f3}
\end{center}
\end{figure}

For a system of $N$ particles with periodic boundary conditions, the total force-tile area $A_N$ of a particular jammed packing is predicted \cite{Tighe3} to be exactly determined by the total system stress $\Gamma_N$, via the relation,
\begin{equation}
A_N =\frac{\Gamma_N^2}{V}=p^2V.
\label{AN}
\end{equation}
As we sample different mechanically stable configurations at fixed $\Gamma_N$,  the total system volume $V$ fluctuates slightly.  Averaging over these different configurations, the above becomes, $\langle A_N\rangle = \Gamma_N^2\left\langle\frac{1}{V}\right\rangle$.
In Fig.~\ref{f3} we plot the resulting $[\langle A_N\rangle /(\Gamma_N^2\left\langle\frac{1}{V}\right\rangle)]-1$ vs the total system stress per particle $\tilde p$.  
From Eq.~(\ref{AN}) we expect this quantity to vanish.  Our numerical results show that $\langle A_N\rangle /(\Gamma_N^2\left\langle\frac{1}{V}\right\rangle)$ is indeed equal to unity within  roughly $2\times 10^{-6}$.  The small discrepancy is presumably due to the failure to achieve perfect force balance in our minimization of $\tilde U$.

Because our total system is sufficiently large, the relative fluctuations in the total system volume $V$ from configuration to configuration are roughly only 0.03-0.04\%.  In the following we will therefore view $V\equiv\langle V\rangle$ and $A_N\equiv\langle A_N\rangle$ as fixed values.

{\em Voronoi Volume}: The Voronoi volume of a particle, $V_i$, is defined as the region of space closer to particle $i$ than to any other particle.  Since every point in space is closest to some particle, the volumes $V_i$ tile all of position space with no gaps or overlaps.  The Voronoi volume of a cluster is just $V_{\cal C}=\sum_{i\in{\cal C}}V_i$, and the sum over all particles is just the total volume of the entire system $V$.  The Voronoi volume is thus a conserved quantity.  We use the Voro++ software package to determine the Voronoi volumes of our particles \cite{voro}.

{\em Number of Particles}: The total number of particles $N_{\cal C}$ in a cluster is clearly also a conserved quantity.  For a bidisperse system, such as we study here, so is the number of small particles $N_{s{\cal C}}$ contained in the cluster.

\section{Clusters with Fixed Radius $R$}

We consider first the clusters with a fixed radius $R$, and fluctuating number of particles $N_R$.

\subsection{Averages}

Because of the additive nature of the conserved quantities, we expect that the average value of such a quantity $X$ defined on a cluster will be related to the fixed value of the entire $N$ particle system $X_N$ according to the fraction of the total system occupied by the cluster.  For the quantity $X_R$ defined on clusters of radius $R$ we therefore expect
\begin{equation}
\langle X_R\rangle=X_N\left(\frac{\pi R^2}{V}\right)\>\Rightarrow\>\left( \frac{\langle X_R\rangle}{\pi R^2}\right)\left(\frac{V}{X_N}\right)=1.
\label{XR}
\end{equation}
If Eq.~(\ref{XR}) holds, then knowledge of the global system parameters gives knowledge about the expected values on clusters within the system.  This is what makes such quantities useful for formulating a maximum entropy model of fluctuating quantities on subsets of the total system.

\begin{figure}[h!]
\begin{center}
\includegraphics[width=2.8in]{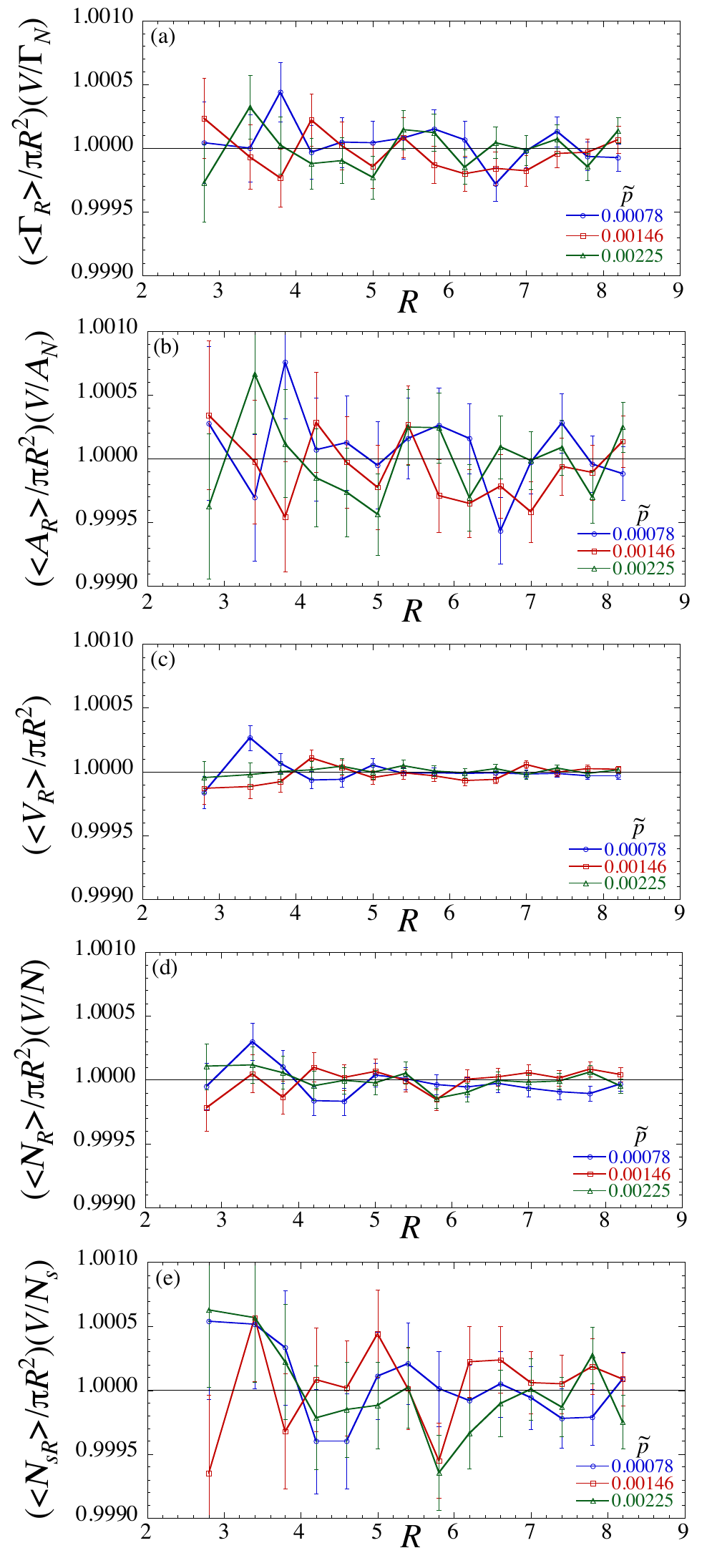}
\caption{(color online) Ratio of intensive quantities defined on a cluster of radius $R$, $(X_R/\pi R^2)$, to the corresponding quantity defined on the total system, $(X_N/V)$, vs cluster radius $R$ for $X$ equal to the  (a) stress $\Gamma$; (b) force-tile area $A$; (c) Voronoi volume $V$; (d) number of particles $N$; (e) number of small particles $N_s$. Three different values of the total system stress per particle $\tilde p$ are shown, represented by three different symbol shapes. Our system has a total of $N=8192$ particles.
}
\label{f4}
\end{center}
\end{figure}

In Fig.~\ref{f4} we plot $(\langle X_R\rangle/\pi R^2)(V/X_N)$ vs $R$ for $X_R=\Gamma_R,A_R,V_R,N_R$ and $N_{sR}$, at three different values of the total system stress per particle $\tilde p=\Gamma_N/N$.  We see that the deviations from the expected value of unity are very small (less than 
$0.1\%$ for $\Gamma_R$, $A_R$, and $N_{sR}$, and less than 0.02\% for $V_R$ and $N_R$) and are all within the estimated statistical error. 
In particular, these results confirm that the average Voronoi volume of the cluster is just the area of the circle, $\langle V_R\rangle = \pi R^2$.

\subsection{Variances}

\begin{figure}[h!]
\begin{center}
\includegraphics[width=3in]{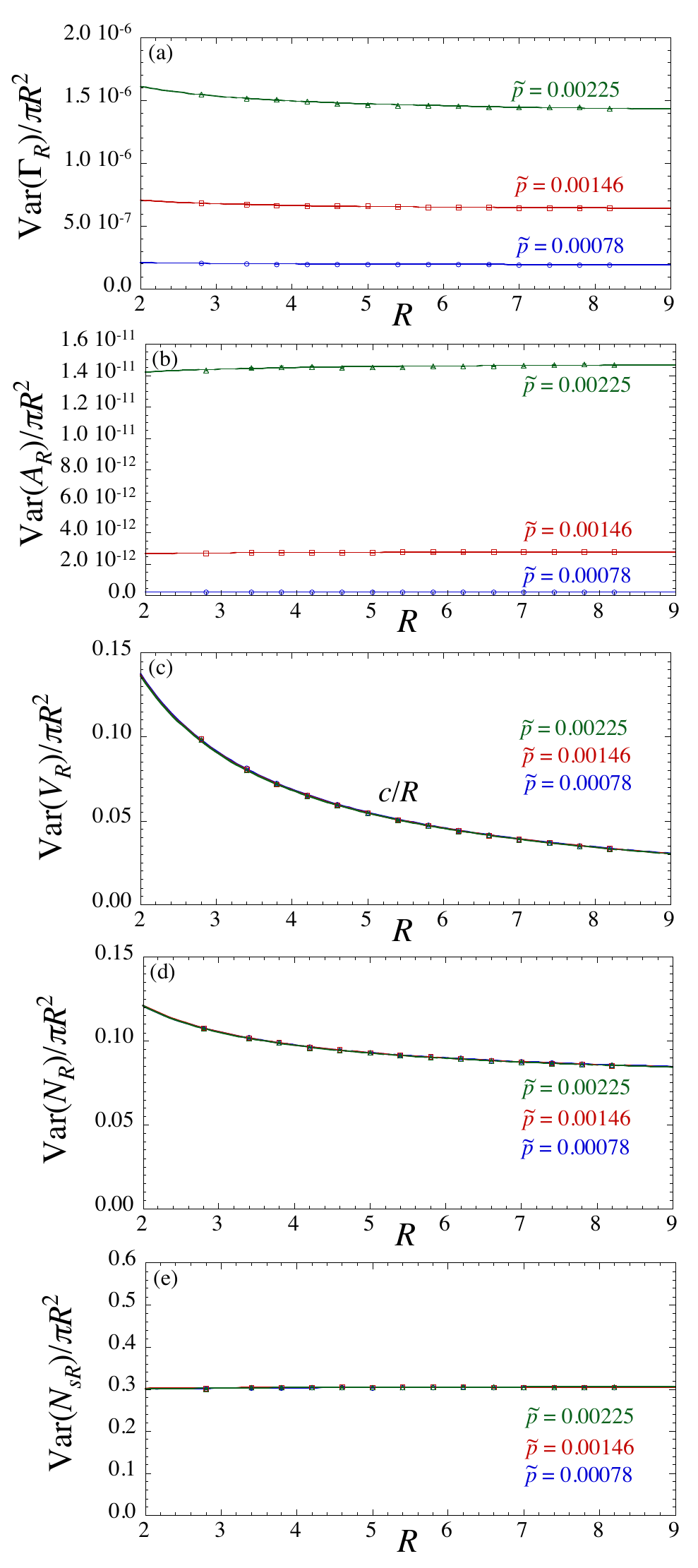}
\caption{(color online) Variance of quantities defined on a cluster of radius $R$, $\mathrm{Var}(X_R)/\pi R^2$ vs $R$, for $X$ equal to the (a) stress $\Gamma$; (b) force-tile area $A$; (c) Voronoi volume $V$; (d) number of particles $N$; (e) number of small particles $N_s$. Three different values of the total system stress per particle $\tilde p$ are shown, represented by three different symbol shapes.  Solid lines are fits to the form $c_1+c_2/R$; for $V_R$, $c_1=0$.
Our system has a total of $N=8192$ particles.
}
\label{f5}
\end{center}
\end{figure}

Now we consider the fluctuations away from the average, and compute the variances of the conserved quantities, $\mathrm{Var}(X_R)\equiv \langle X_R^2\rangle-\langle X_R\rangle^2$.
In Fig.~\ref{f5} we plot $\mathrm{Var}(X_R)/\pi R^2$, vs $R$ for $X_R=\Gamma_R,A_R,V_R,N_R$, and $N_{sR}$, at three different values of the total system stress per particle $\tilde p$.  The solid lines in Figs.~\ref{f5} are fits to the form $c_1+c_2/R$.

For $\Gamma_R$, $A_R$, $N_R$ and $N_{sR}$ we find that the scaled variances all approach a finite constant as $R$ increases, i.e. $c_1>0$.  Thus the variances of these quantities scale proportional to the cluster volume.  
Since, from Fig.~\ref{f4}, the averages of these quantities also scale proportional to the volume, we conclude that their relative fluctuations decay as, 
\begin{equation}
\dfrac{\sqrt{\mathrm{Var}(X_R)}}{\langle X_R\rangle}\propto\dfrac{1}{R}\propto \dfrac{1}{\sqrt{\langle N_R\rangle}}.  
\label{VarXR}
\end{equation}
Such behavior, resulting from the variances being extensive quantities, is just what one would expect if the cluster variable $X_R$ was the sum of independent random variables $X_k$ representing the value of $X$ on subunits of the cluster.  This result therefore suggests that there are, on average, no spatial correlations of $X_R$ on length scales larger than our smallest value of $R=2.8$ \cite{Lois}.

The Voronoi volume $V_R$, however, behaves differently.  The solid line in Fig.~\ref{f5}c is a fit to $c_2/R$ (i.e. taking $c_1=0$), showing that the variance of $V_R$ scales proportional to the {\em perimeter} $\sim R$ of the cluster, rather than its volume.  This is reasonable as only changes in the positions of the particles at the surface of the cluster will effect the Voronoi volume \cite{note1}.  The relative fluctuations of $V_R$ therefore scale as 
\begin{equation}
\dfrac{\sqrt{\mathrm{Var}(V_R)}}{\langle V_R\rangle}\propto\dfrac{ 1}{R^{3/2}}\propto \dfrac{1}{\langle N_R\rangle^{3/4}}.
\label{VR}
\end{equation}

\begin{figure}[h!]
\begin{center}
\includegraphics[width=3in]{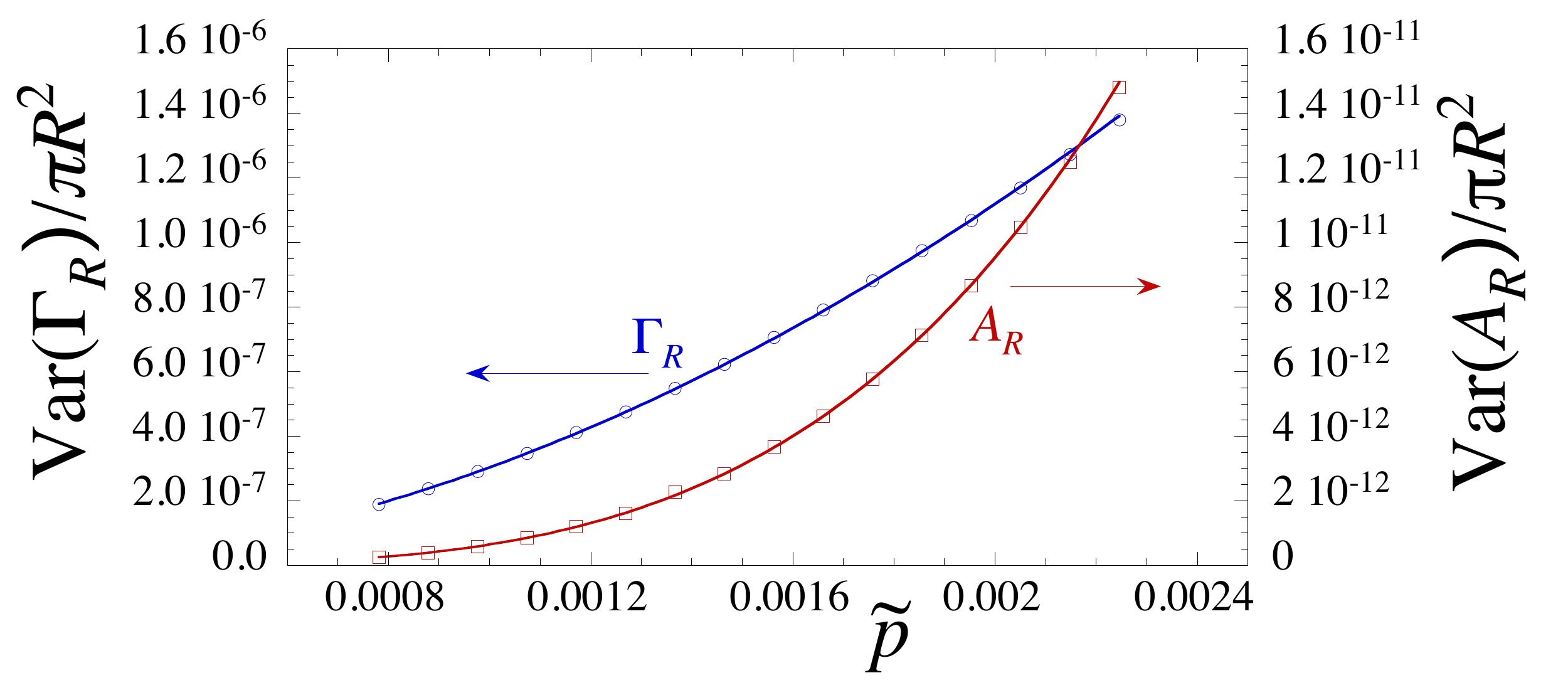}
\caption{(color online) $\mathrm{Var}(\Gamma_R)/\pi R^2$ and $\mathrm{Var}(A_R)/\pi R^2$, in the large $R$ limit, vs total stress per particle $\tilde p$.  Solid lines are fits to power laws and give $\sim \tilde p^{1.9}$ and $\sim \tilde p^{3.9}$, respectively. Our system has a total of $N=8192$ particles.
}
\label{f6}
\end{center}
\end{figure}

From Fig.~\ref{f5} see that only the scaled variances of $\Gamma_R$ and $A_R$ show a dependence on $\tilde p$, with increasing fluctuations as $\tilde p$ increases.  
In Fig.~\ref{f6} we plot the large $R$ limiting value of $\mathrm{Var}(X_R)/\pi R^2$ for these two quantities, vs  $\tilde p$.  The solid lines in Fig.~\ref{f6} are power law fits; we find $\mathrm{Var}(\Gamma_R)/\pi R^2\sim \tilde p^{1.9}$, while $\mathrm{Var}(A_R)/\pi R^2\sim \tilde p^{3.9}$ \cite{note2}.  Since by Eqs.~(\ref{AN}-\ref{XR})  $\langle\Gamma_R\rangle/\pi R^2=\Gamma_N/V=\tilde p (N/V)$, and $\langle A_R\rangle/\pi R^2=A_N/V=(\Gamma_N/V)^2=\tilde p^2 (N/V)^2$, we conclude that the relative fluctuations, $\sqrt{\mathrm{Var}(\Gamma_R)}/\langle\Gamma_R\rangle$ and $\sqrt{\mathrm{Var}(A_R)}/\langle A_R\rangle$ both scale as $c/R$, with a constant $c$ that is only weakly dependent on the total stress per particle $\tilde p$.

\subsection{Hyperuniformity}

\begin{figure}[h!]
\begin{center}
\includegraphics[width=3in]{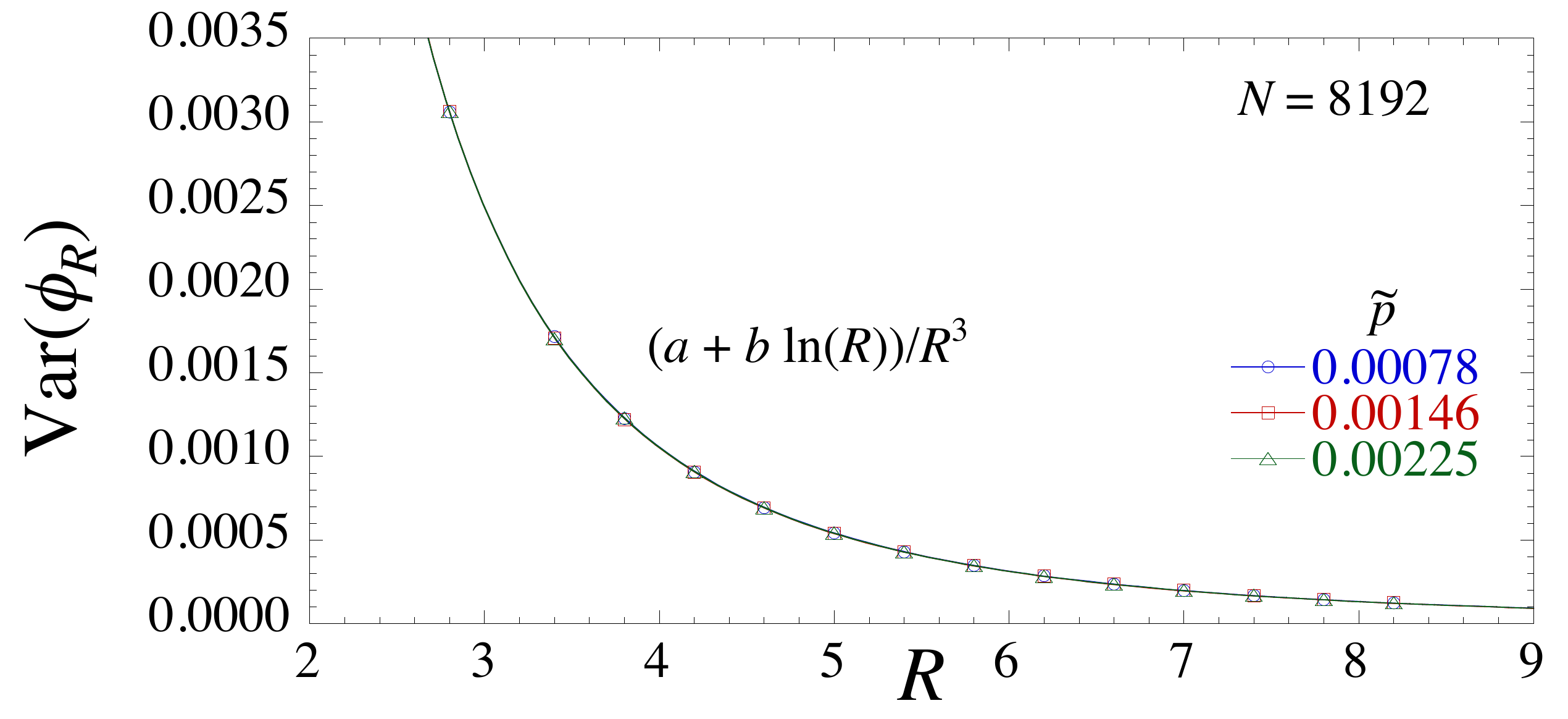}
\caption{(color online) $\mathrm{Var}(\phi_R)$ vs $R$, for three values of the total stress per particle $\tilde p$, represented by three different symbol shapes.  Solid lines are fits to $(a+b\ln(R))/R^3$. Our system has a total of $N=8192$ particles.
}
\label{f7}
\end{center}
\end{figure}

It is interesting to consider the variance of the local average packing fraction,
\begin{equation}
\langle \phi_R\rangle = \frac{v_b\langle N_{bR}\rangle  + v_s\langle N_{sR}\rangle }{\pi R^2},
\end{equation}
where $N_{bR}=N_R-N_{sR}$ is the number of big particles in the cluster.  We have,
\begin{equation}
\begin{aligned}
\mathrm{Var}(\phi_R)=&\left(v_b^2\mathrm{Var}(N_{bR})+v_s^2\mathrm{Var}(N_{sR}) \right.\\[10pt]
&\left.+2 v_bv_s\mathrm{Covar}(N_{bR},N_{sR})\right)/({\pi R^2})^2
\end{aligned}
\end{equation}
where the covariance $\mathrm{Covar}(N_{bR},N_{sR})\equiv \langle N_{bR}N_{sR}\rangle - \langle N_{bR}\rangle\langle N_{sR}\rangle$.  Since the variances above, and also (as we have checked) the covariance, scale proportional to $\pi R^2$, one would naively expect $\mathrm{Var}(\phi_R)\propto 1/R^2$.  In Fig.~\ref{f7} we plot $\mathrm{Var}(\phi_R)$ vs $R$, for three different values of $\tilde p$.  We see that $\mathrm{Var}(\phi_R)$ is independent of $\tilde p$, and in contrast to the naive expectation vanishes more rapidly, like $\sim 1/R^3$.  This is a signature of the {\em hyperuniformity} of jammed packings that has been observed exactly at $\phi_J$ by earlier analyses of the long wavelength limit of the compressibility \cite{Sollich} and the 2-point spectral density function \cite{Zachary}.  The solid line in Fig.~\ref{f7} is a fit to the form $(a+b\ln(R))/R^3$, as predicted by Ref.~\cite{Zachary}.  Although the inclusion of the $\ln(R)$ term in this formula does slightly improve the fit, our range of $R$ is too limited to provide a sensitive test for the existence of this logarithmic term; the overall $1/R^3$ dependence, however, is unmistakable. Our results show that the hyperuniformity, previous observed at $\phi_J$, persists in packings above $\phi_J$, at least on the length scales we are able to investigate (we cannot rule out a change in behavior for much larger $R$).

\subsection{Correlations Between Conserved Quantities}

Finally we consider the correlations between the conserved quantities.  Since the stress and the Voronoi volume have been the variables previously used to construct maximum entropy models of the statistics of jammed packings, we focus here on the correlations between $\Gamma_R$ and the other conserved variables, and then on the correlations between $V_R$ and the remaining conserved variables.
To compare quantities on similar scales, we consider here the rescaled variables, 
\begin{equation}
\hat X_R\equiv \dfrac{(X_R-\langle X_R\rangle)}{\sigma_{X_R}}, 
\label{hatXR}
\end{equation}
where $\sigma_{X_R}$ is the standard deviation of $X_R$.
To highlight the quadratic relation between $\Gamma_R$ and $A_R$, instead of $A_R$ we consider here $A_R^{1/2}$, which is linearly related to $\Gamma_R$. 

\begin{figure}[h]
\begin{center}
\includegraphics[width=2.8in]{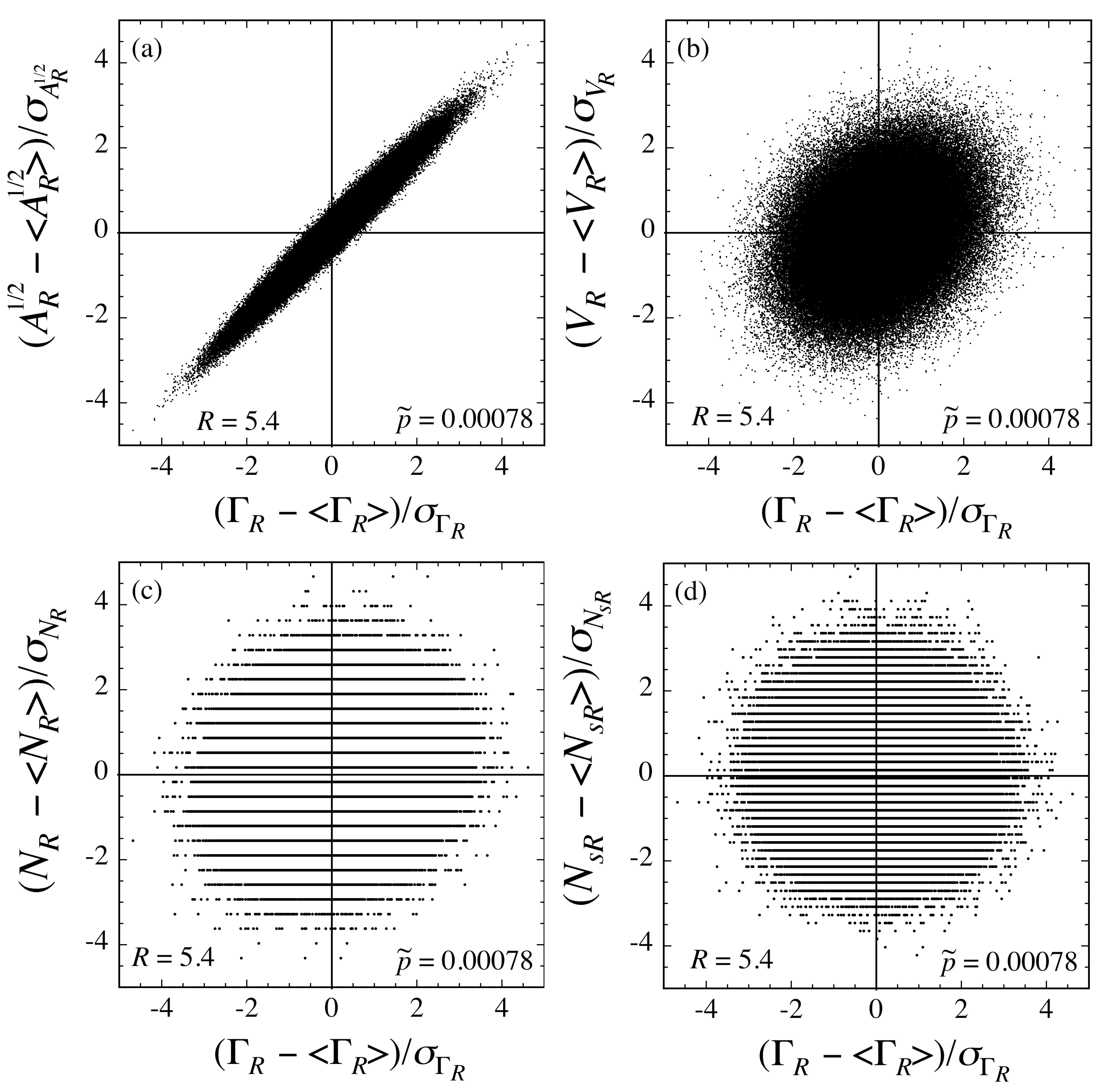}
\caption{Scatter plots showing configuration specific values of $(\Gamma_R-\langle\Gamma_R\rangle)/\sigma_{\Gamma_R}$ vs $(X_R-\langle X_R\rangle)/\sigma_{X_R}$ for $X_R$ equal to the (a) square root of the force-tile area $A_R^{1/2}$; (b) Voronoi volume $V_R$; (c) number of particles $N_R$; (d) number of small particles $N_{sR}$.  Here $\sigma_{X_R}$ is the standard deviation of variable $X_R$ and results are for the specific case $R=5.4$ and $\tilde p=0.00078$.
}
\label{f8}
\end{center}
\end{figure}

\begin{figure}[h]
\begin{center}
\includegraphics[width=2.8in]{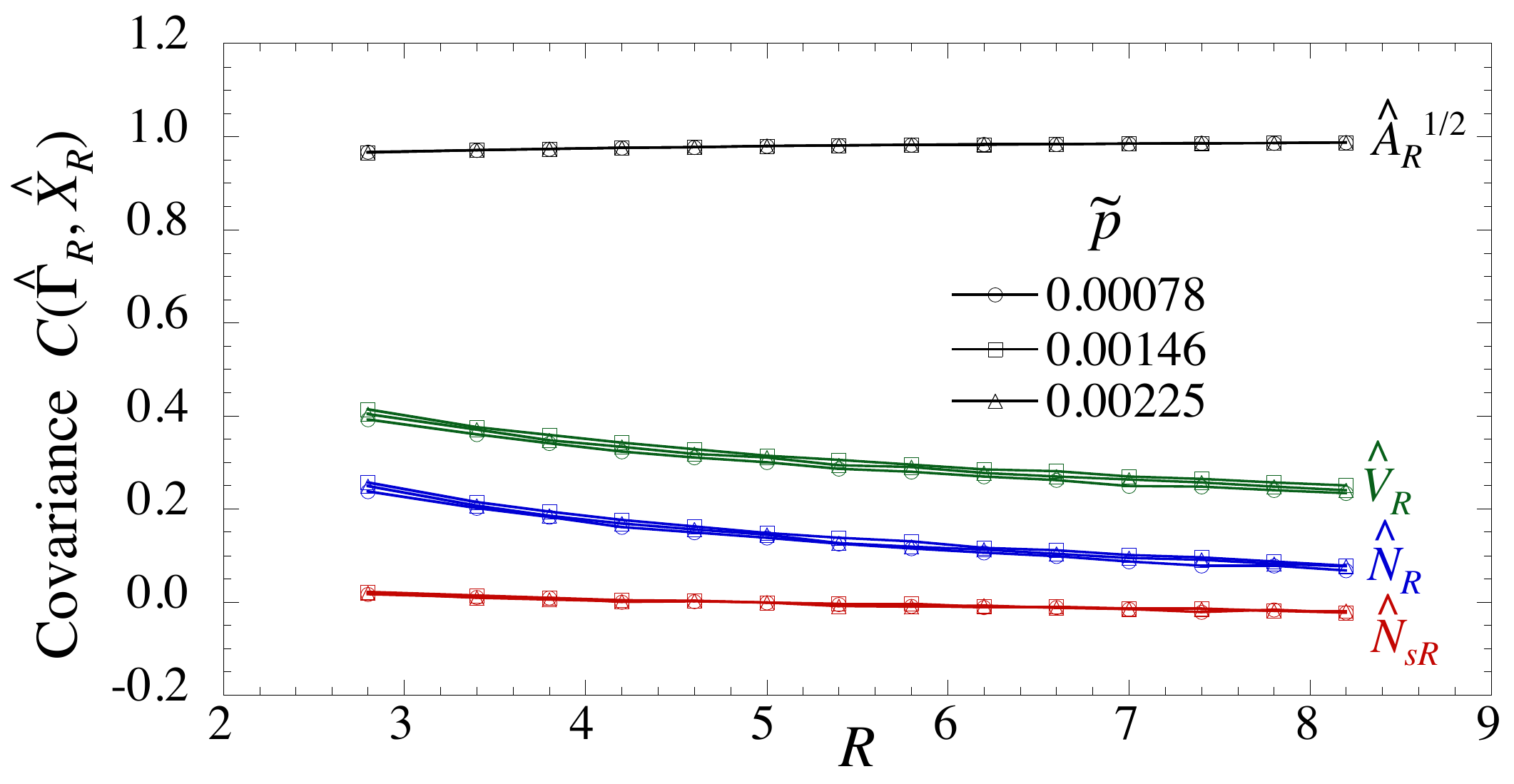}
\caption{(color online) Covariance between rescaled stress $\hat\Gamma_R$ and other variables vs cluster radius $R$, for three different values of total system stress per particle $\tilde p$.  The rescaled variables are defined by $\hat X_R\equiv (X_R-\langle X_R\rangle)/\sigma_{X_R}$, with $\sigma_{X_R}$ the standard deviation of $X_R$, and the plot shows results for $X_R$ equal to the square root of the force-tile area $A_R^{1/2}$, Voronoi volume $V_R$, number of particles $N_R$, and number of small particles $N_{sR}$. Our system has a total of $N=8192$ particles.
}
\label{f9}
\end{center}
\end{figure}

We consider first the correlations with the stress $\Gamma_R$.  In Fig.~\ref{f8} we show scatter plots of the configuration specific values of $\hat \Gamma_R$ vs the other variables, for the particular case of $R=5.4$ and $\tilde p=0.00078$. 
From the scatter plots we see qualitatively the very strong linear correlation between $\hat \Gamma_R$ and $\hat A_R^{1/2}$.  The correlation between $\hat \Gamma_R$ and $\hat V_R$ is in comparison considerably weaker, the correlation between $\hat \Gamma_R$ and $\hat N_R$ is even weaker, and the correlation between $\hat \Gamma_R$ and $\hat N_{sR}$ is almost non existent.  
To quantify this, we plot in Fig.~\ref{f9} the covariances between $\hat\Gamma_R$ and the other variables vs cluster radius $R$, for three different values of the total system stress per particle $\tilde p$.  We see that the covariances are essentially independent of $\tilde p$.  The covariance of $\hat\Gamma_R$ with $\hat A_R^{1/2}$ is very large, increasing towards the maximum value of unity as $R$ increases.  The next strongest correlation is with the Voronoi volume $\hat V_R$, then the number of particles $\hat N$.  The covariance with the number of small particles $\hat N_s$ is very small, about 1\%.  Moreover, we see that the covariance of $\hat\Gamma_R$ with $\hat V_R$ and $\hat N_R$ is decreasing as $R$ increases.  We do not have a large enough range of $R$ to determine whether these correlations vanish as $R\to\infty$, or saturate to a finite value.  

Although the correlation of $\hat\Gamma_R$ with $\hat V_R$, at 20\% for our largest $R$, is perhaps not insignificant, recall that this is the correlation of the rescaled variables.  If we consider instead the correlation of the relative fluctuations, $\left[\mathrm{Covar}(\Gamma_R,V_R)/(\langle\Gamma_R\rangle\langle V_R\rangle)\right]^{1/2}$, the suppressed relative fluctuations of $V_R$ given by Eq.~(\ref{VR}) will mean similarly suppressed relative fluctuations of the correlations, which will decay as $\sim 1/R^{5/4}$, faster than the $\sim 1/R$ decay of other relative correlations.

The effect of correlations between stress and Voronoi volume on the statistical description of jammed packings has recently been considered by Blumenfeld et al. \cite{Blumenfeld}, who argue that these correlations preclude the use of either a volume-only, or a stress-only statistical ensemble.  Our result here, that the stress-volume correlation decreases as the cluster size increases, suggests that the effects of such correlations may become less significant on longer length scales, for clusters of fixed radius.

\begin{figure}[h]
\begin{center}
\includegraphics[width=2.8in]{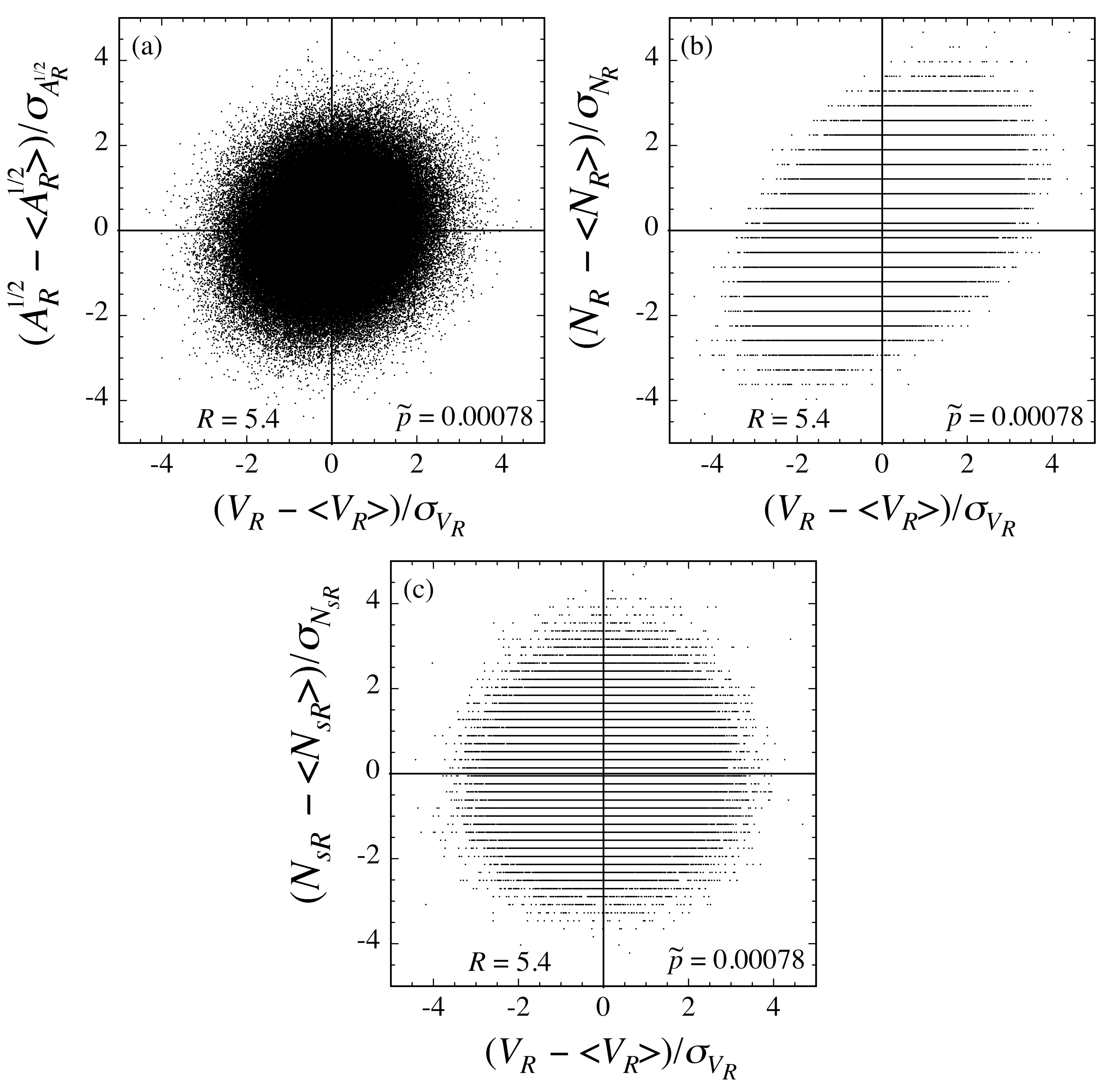}
\caption{Scatter plots showing configuration specific values of $(V_R-\langle V_R\rangle)/\sigma_{V_R}$ vs $(X_R-\langle X_R\rangle)/\sigma_{X_R}$ for $X_R$ equal to the (a) square root of the force-tile area $A_R^{1/2}$;  (b) number of particles $N_R$; (c) number of small particles $N_{sR}$.  Here $\sigma_{X_R}$ is the standard deviation of variable $X_R$ and results are for the specific case $R=5.4$ and $\tilde p=0.00078$.
}
\label{f10}
\end{center}
\end{figure}

\begin{figure}[h]
\begin{center}
\includegraphics[width=2.8in]{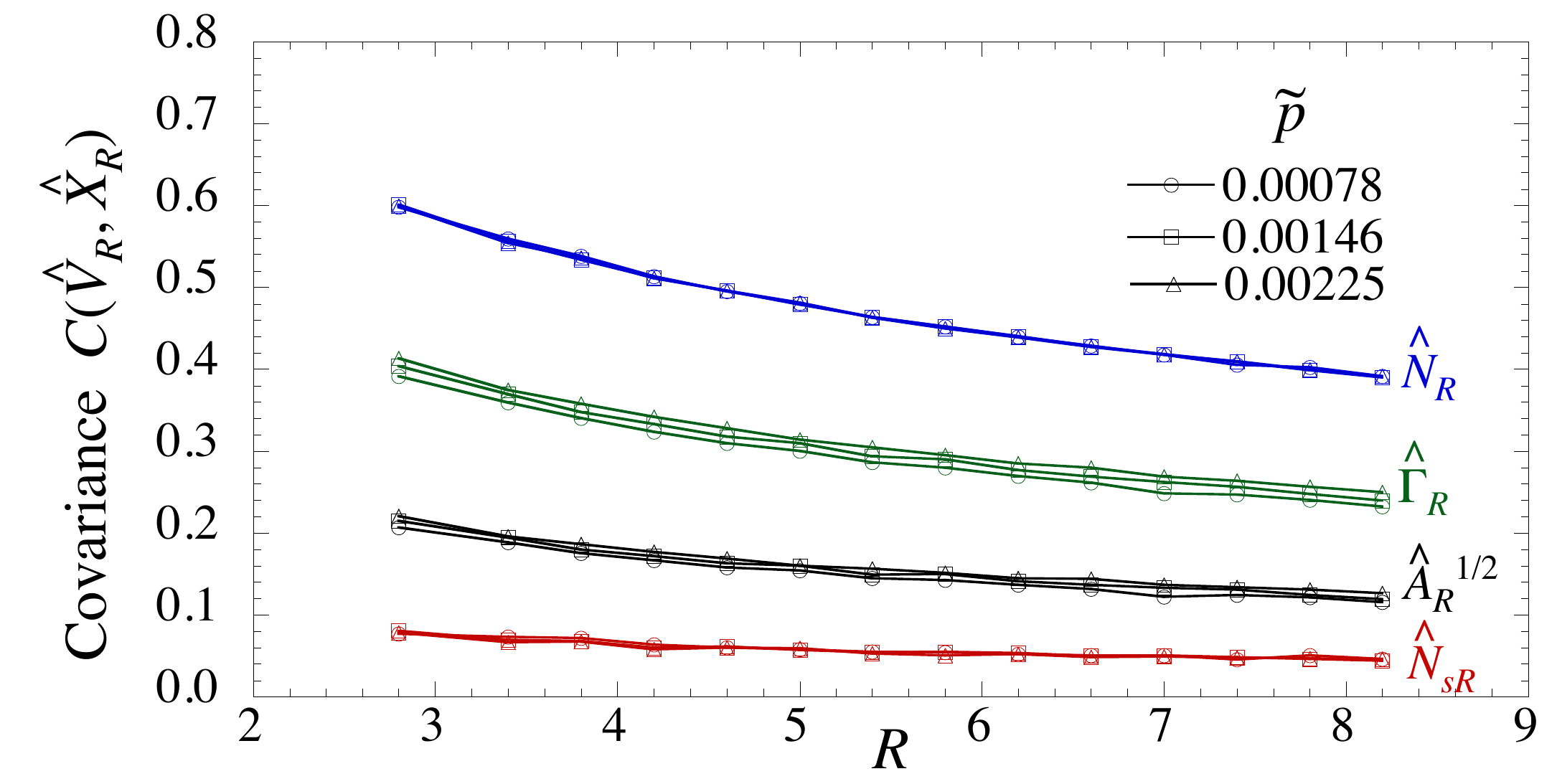}
\caption{(color online) Covariance between rescaled Voronoi volume $\hat V_R$ and other variables vs cluster radius $R$, for three different values of total system stress per particle $\tilde p$.  The rescaled variables are defined by $\hat X_R\equiv (X-\langle X_R\rangle)/\sigma_{X_R}$, with $\sigma_{X_R}$ the standard deviation of $X_R$.  We shows results for $X_R$ equal to the stress $\Gamma_R$,  square root of the force-tile area $A_R^{1/2}$, number of particles $N_R$, and number of small particles $N_{sR}$. Our system has a total of $N=8192$ particles.
}
\label{f11}
\end{center}
\end{figure}

We turn now to the correlations with the Voronoi volume $V_R$.  In Fig.~\ref{f10} we show scatter plots of the configuration specific values of $\hat V_R$ with $\hat A_R^{1/2}$, $N_R$, and $N_{sR}$, for the particular case of $R=5.4$ and $\tilde p=0.00078$. In Fig.~\ref{f11} we plot the covariance between $\hat V_R$ and the other variables vs cluster radius $R$, for three different values of the total system stress per particle $\tilde p$.  We see that the covariances are essentially independent of $\tilde p$. For $\hat V_R$, the strongest correlation is with $\hat N_R$.  Again, we see that the correlations decrease as the cluster size $R$ increases.

\section{Clusters of fixed number of particles $M$}

In this section we consider the second of our two cluster ensembles, clusters which contain a fixed number of particles $M$.  We will see some striking differences between the statistical behavior of these clusters and the previously discussed clusters of fixed radius $R$.  Such fixed $M$ clusters have been used in some earlier numerical works \cite{Henkes1, Henkes2}.

\subsection{Averages}

\begin{figure}[h!]
\begin{center}
\includegraphics[width=2.8in]{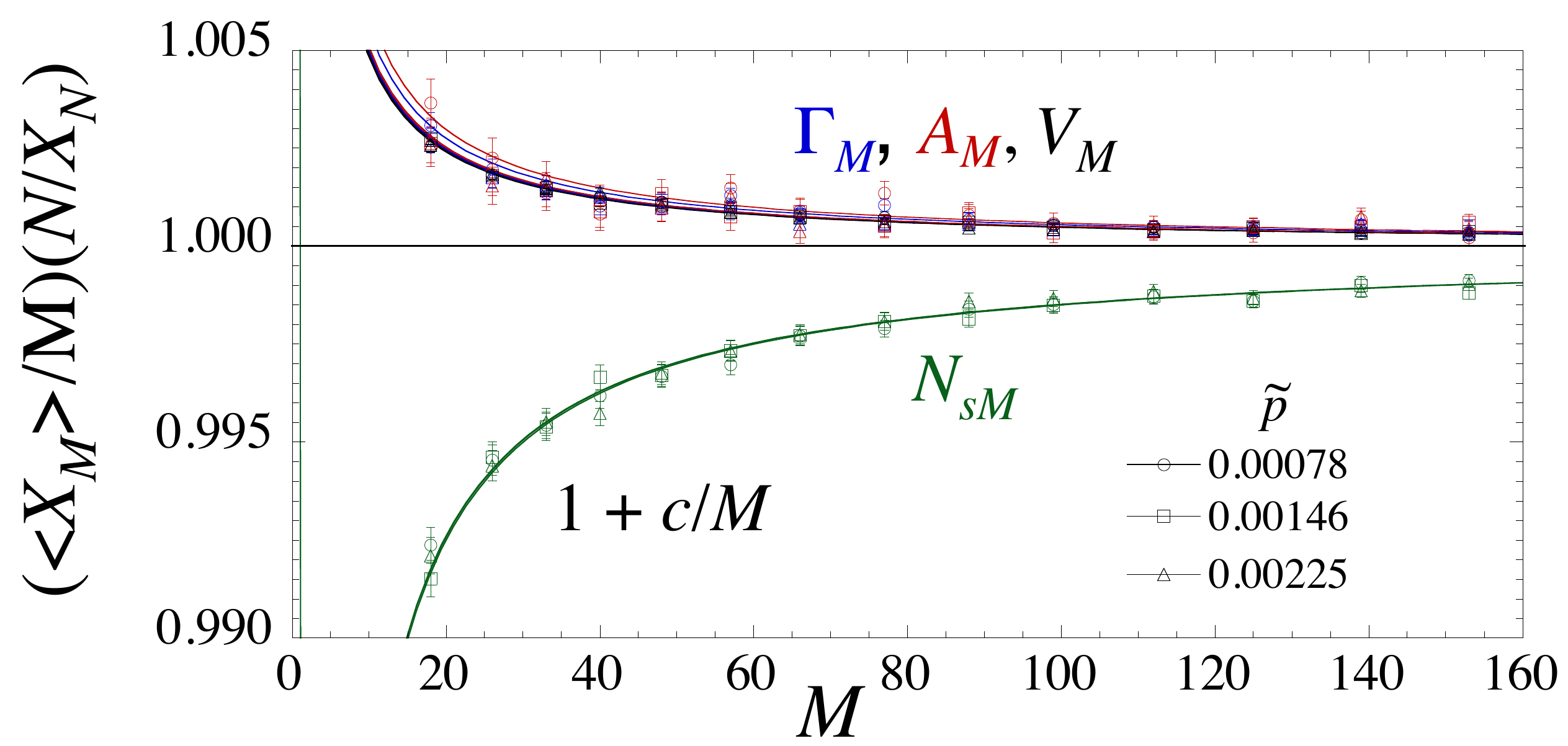}
\caption{(color online) Ratio of intensive quantities defined on a cluster of fixed number of particles $M$, $(X_M/M)$, to the corresponding quantity defined on the total system, $(X_N/N)$, vs $M$ for $X$ equal to the stress $\Gamma$, force-tile area $A$, Voronoi volume $V$, and number of small particles $N_s$. Three different values of the total system stress per particle $\tilde p$ are shown, represented by three different symbol shapes.  Solid lines are fits to the form $1+c/M$. Our system has a total of $N=8192$ particles.
}
\label{f12}
\end{center}
\end{figure}

One might expect that averages of conserved quantities on such clusters $X_M$ would just be equal to the fraction of total particles $(M/N)$ contained in the cluster times the corresponding total system quantity, $X_N$.  In Fig.~\ref{f12} we plot $(X_M/M)(N/X_N)$ vs $M$, for $X_M=\Gamma_M,A_M,V_M$, and $N_{sM}$, for three different values of the total system stress per particle $\tilde p$.  However, in contrast to the corresponding quantities defined for clusters of fixed $R$ which are equal to unity at all $R$ (see Fig.~\ref{f4}), here we find that these quantities only approach unity algebraically as $M$ increases. The solid lines in Fig.~\ref{f12} are fits to the form $1+c/M$.  The behavior of $(X_M/M)(N/X_N)$ is independent of $\tilde p$, and appears to be identical, decreasing towards unity, for $X_M=\Gamma_M,A_M$, and $V_M$; for $N_{sM}$ the effect is about double, and has the opposite sign, increasing towards unity as $M$ increases.  Thus, for clusters of fixed number of particles, unlike the clusters of fixed radius $R$, the averages of the conserved quantities are {\em not} simply fixed by the global average value $X_N/N$, but rather depend on the cluster size $M$ in a way that is not apriori known (i.e. the coefficient $c$ must be determined from other information). 

\subsection{Variances}

\begin{figure}
\begin{center}
\includegraphics[width=3in]{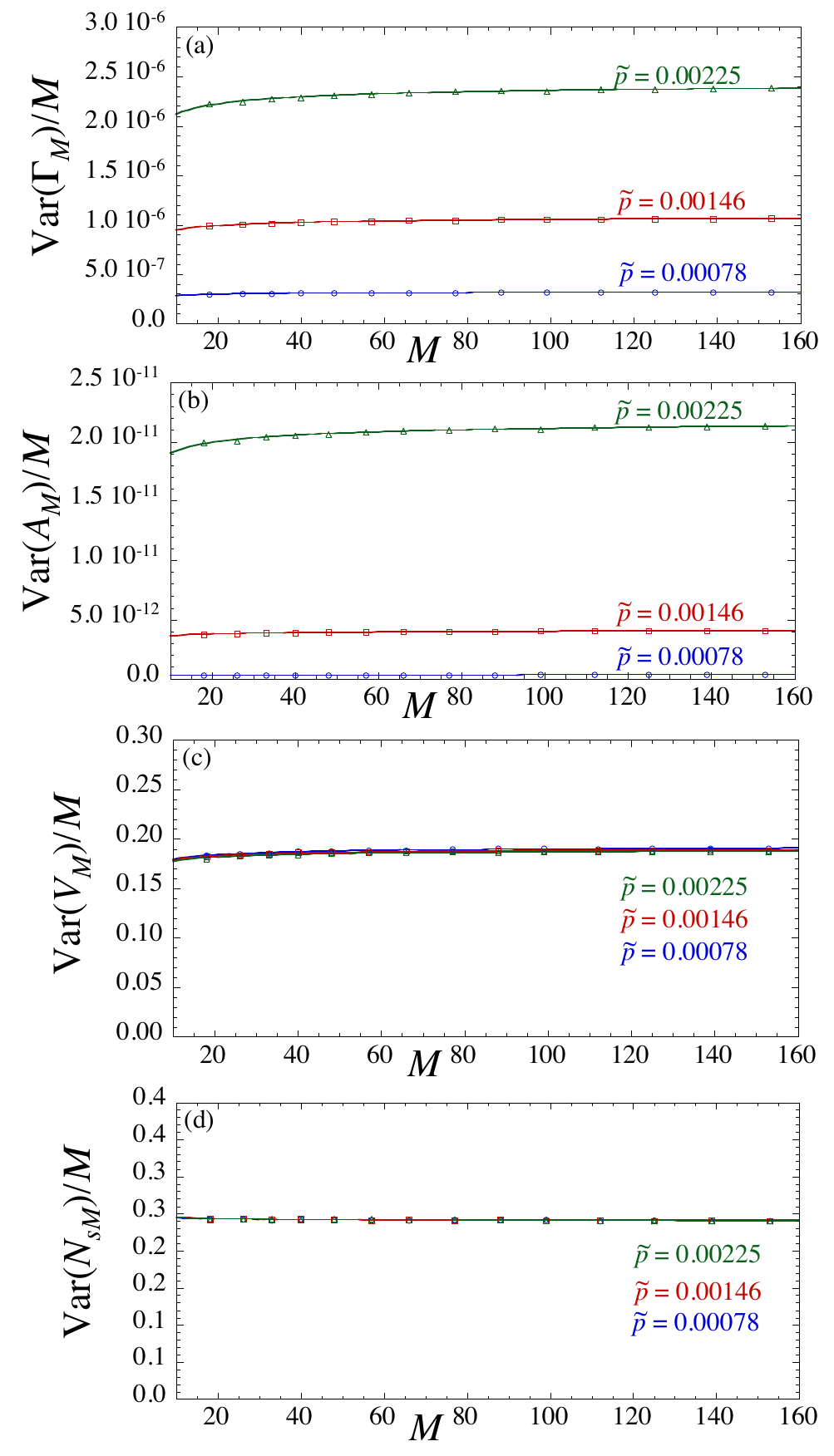}
\caption{(color online) Variance of quantities defined on a cluster of fixed number of particles $M$, $\mathrm{Var}(X_M)/M$ vs $M$, for $X$ equal to the (a) stress $\Gamma$; (b) force-tile area $A$; (c) Voronoi volume $V$; (d) number of small particles $N_s$. Three different values of the total system stress per particle $\tilde p$ are shown, represented by three different symbol shapes.  Solid lines are fits to the form $c_1+c_2/\sqrt{M}$.
Our system has a total of $N=8192$ particles.
}
\label{f13}
\end{center}
\end{figure}

We can also look at the variance of the conserved quantities on clusters of fixed number of particles $M$.  In Fig.~\ref{f13} we plot $\mathrm{Var}(X_M)/M$ vs $M$ for $X_M=\Gamma_M, A_M, V_M$, and $N_{sM}$.  We show results for three different values of $\tilde p$.  The results are qualitatively similarly to what was seen in Fig.~\ref{f5} for the clusters of fixed radius $R$. Only the behavior of the Voronoi volume $V_R$ is different; instead of $\mathrm{Var}(V_R)\sim R$ growing as the perimeter of the cluster, we now have $\mathrm{Var}(V_M)\sim M$, i.e. growing proportional to the cluster volume, just like the other quantities.

\begin{figure}[h!]
\begin{center}
\includegraphics[width=3in]{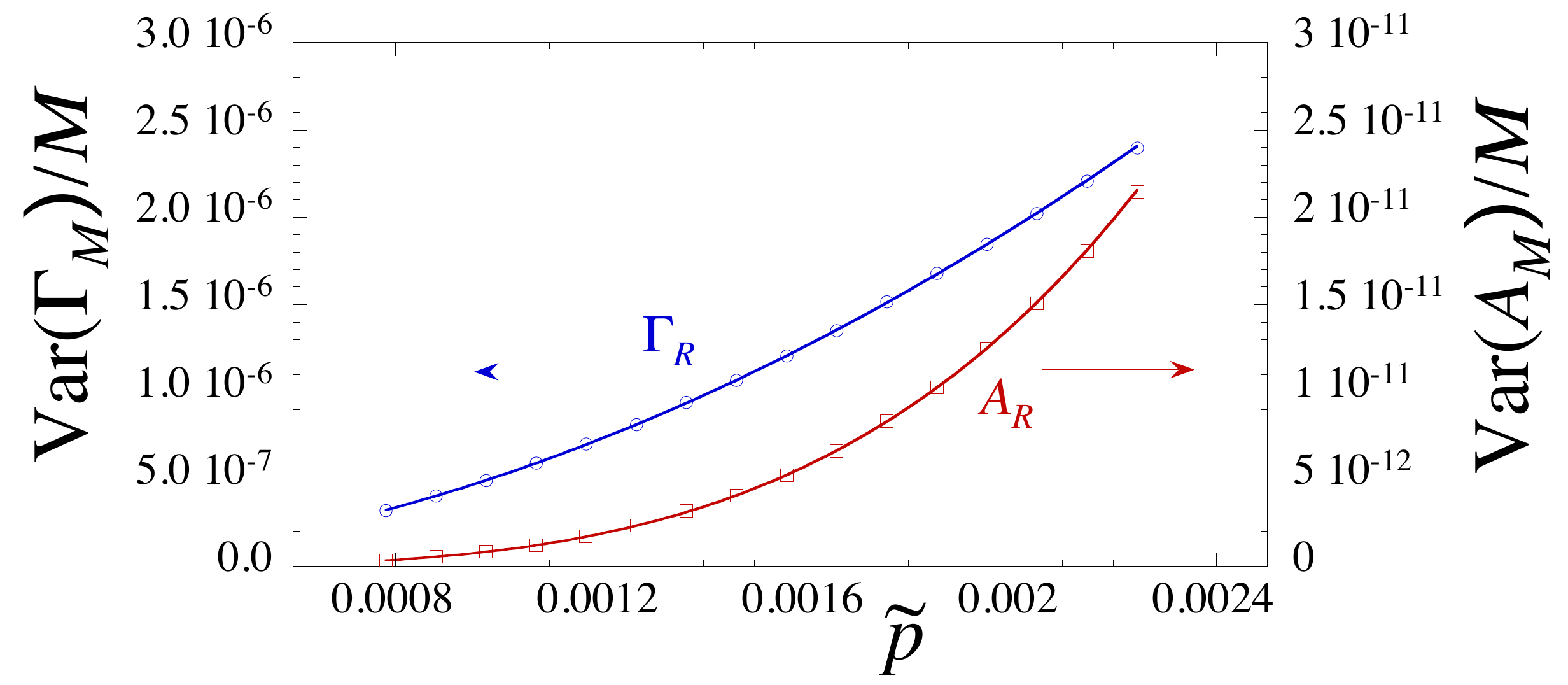}
\caption{(color online) $\mathrm{Var}(\Gamma_M)/M$ and $\mathrm{Var}(A_M)/M$, in the large $M$ limit, vs total stress per particle $\tilde p$.  Solid lines are fits to power laws and give $\sim \tilde p^{1.9}$ and $\sim \tilde p^{3.9}$, respectively. Our system has a total of $N=8192$ particles.
}
\label{f14}
\end{center}
\end{figure}

We again find that only $\mathrm{Var}(\Gamma_M)/M$ and $\mathrm{Var}(A_M)/M$ vary significantly with the  total stress per particle $\tilde p$.  Plotting the large $R$ limiting value of $\mathrm{Var}(\Gamma_M)/M$ and $\mathrm{Var}(A_M)/M$ vs $\tilde p$ in Fig.~\ref{f14}, we find that they have the same behavior as was found previously for clusters of fixed radius $R$, $\mathrm{Var}(\Gamma_M)/M\propto \tilde p^{1.9}$ and $\mathrm{Var}(A_M)/M\propto \tilde p^{3.9}$ \cite{note2}.

\subsection{Concentration of Small particles}

Our observation in Fig.~\ref{f12} implies that the average concentration of small particles in a cluster, $x_s(M)\equiv\langle N_{sM}\rangle/M$, is not fixed at the global value $N_s/N=1/2$, but varies algebraically with the cluster size, approaching 1/2 only as the cluster grows large.  We believe it is this that effects the dependence of all the other variables on $M$.  

For clusters of fixed radius $R$, our results in Fig.~\ref{f4} imply that the ratio $\bar x_s\equiv\langle N_{sR}\rangle/\langle N_R\rangle=1/2$ for all cluster sizes $R$.  However, the average concentration of small particles in such clusters is more properly computed as, 
\begin{equation}
x_s(R)=\left\langle\dfrac{N_{sR}}{N_R}\right\rangle.
\label{xR}
\end{equation}
We can now show that $x_s(R)$ has the same algebraic behavior as $x_s(M)$.
Defining $\delta N_{sR}$ so that $N_{sR}=\langle N_{sR}\rangle +\delta N_{sR}$, with $\langle \delta N_{sR}\rangle = 0$ and $\langle (\delta N_{sR})^2\rangle = \mathrm{Var}(N_{sR})$, and similarly defining $\delta N_R$, we can write,
\begin{equation}
 x_s(R)=\dfrac{\langle N_{sR}\rangle}{\langle N_R\rangle}\left\langle\dfrac{1+\delta N_{sR}/\langle N_{sR}\rangle}{1+\delta N_R/\langle N_R\rangle}\right\rangle.
 \end{equation}
 Expanding for small $\delta N_R$ and $\delta N_{sR}$, we get to second order,
 \begin{equation}
 x_s(R)=\bar x_s\left( 1+\dfrac{\langle\delta N_R^2\rangle}{\langle N_R\rangle^2}-
 \dfrac{\langle\delta N_{sR}\delta N_R\rangle}{\langle N_{sR}\rangle\langle N_R\rangle}\right).
 \end{equation}
 Writing $\langle N_{sR}\rangle =\bar x_s\langle N_R\rangle$ and $N_{bR}\equiv N_R-N_{sR}$, for our particular case of $\bar x_s=1/2$ we get,
 \begin{equation}
 \begin{aligned}
 x_s(R)&=\frac{1}{2}\left(1+\dfrac{\langle\delta N_{bR}^2\rangle-\langle\delta N_{sR}^2\rangle}{\langle N_R\rangle^2}\right)\\[12pt]
 &=\frac{1}{2}\left(1+\dfrac{\mathrm{Var}(N_{bR})-\mathrm{Var}(N_{sR})}{\langle N_R\rangle^2}\right).
 \label{xR2}
 \end{aligned}
 \end{equation}
 Now using the observation that $\mathrm{Var}(N_{sR})$ and $\mathrm{Var}(N_{bR})$ both scale proportional to the cluster volume $\pi R^2$, and that $\langle N_R\rangle$ does as well, we conclude that,
 \begin{equation}
 x_s(R)=\frac{1}{2}\left(1+\frac{\bar c}{R^2}\right)=\frac{1}{2}\left(1+\frac{c}{\langle N_R\rangle}\right),
 \end{equation}
thus showing the same algebraic dependence on the average number of particles $\langle N_R\rangle$ in the cluster as was found in the clusters with fixed number of particles $M$.
 
\begin{figure}[h!]
\begin{center}
\includegraphics[width=3in]{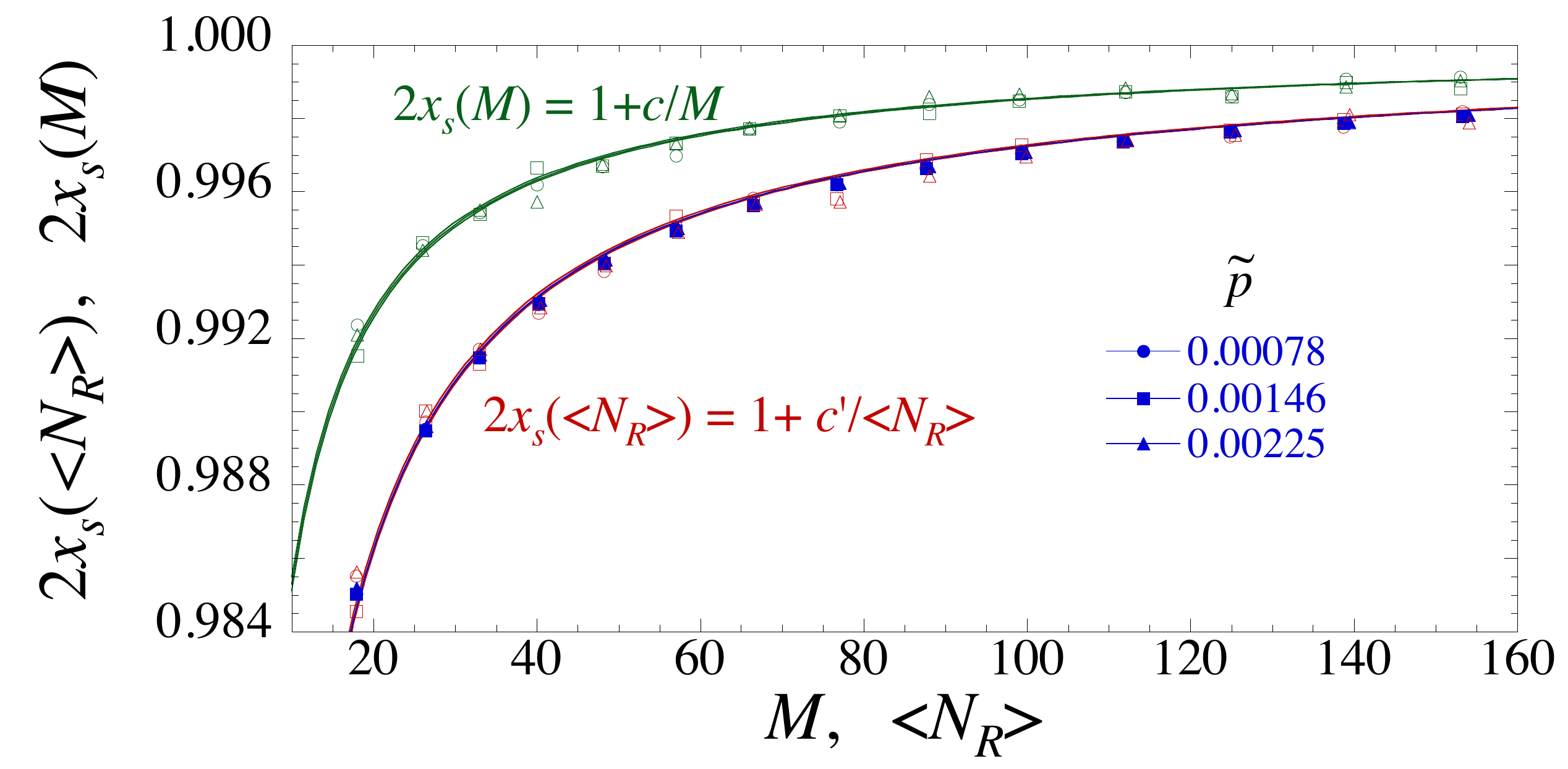}
\caption{(color online) Concentration of small particles $x_s(\langle N_R\rangle)$ and $x_s(M)$ for clusters of fixed radius $R$ and fixed number of particles $M$, respectively.  In the first case, $x_s(\langle N_R\rangle)$ is plotted vs $\langle N_R\rangle$, the average number of particles in the cluster. For $x_s(\langle N_R\rangle)$ we show results from the direct computation of Eq.~(\ref{xR}) (open symbols) as well as the prediction of Eq.~(\ref{xR2}) (solid symbols).  Solid lines are fits to the forms shown.
Three different values of the total system stress per particle $\tilde p$ are shown, represented by three different symbol shapes.
}
\label{f15}
\end{center}
\end{figure}

In Fig.~\ref{f15} we plot $2x_s(R)$ vs $\langle N_{R}\rangle$, for clusters with fixed radius $R$, for three different values of $\tilde p$. We compare the values from a direct computation of $x_s(R)$ from Eq.~(\ref{xR}) (open symbols) with the prediction of Eq.~(\ref{xR2}) (solid symbols) and find excellent agreement.  There is no dependence on $\tilde p$.  In the same figure we also show
$2\langle N_{sM}\rangle/M$ vs $M$ for the clusters with fixed number of particles $M$.  Both show a decay to the large cluster limit of unity that is proportional to the inverse number of particles, however the results are quantitatively somewhat different, presumably due to the different effects of fluctuations in the two ensembles.

We note that if the particles were positioned purely at random, then the concentration $x_s(R)$ of small particles within a circle of radius $R$ would be uniform and equal to the global concentration $\bar x_s$ for any $R$.  The algebraic variation with $R$ that we find here is therefore a consequence of the structural details of how the particles are arranged in the jammed packing.

\subsection{Correlations Between Conserved Quantities}

\begin{figure}[h]
\begin{center}
\includegraphics[width=2.8in]{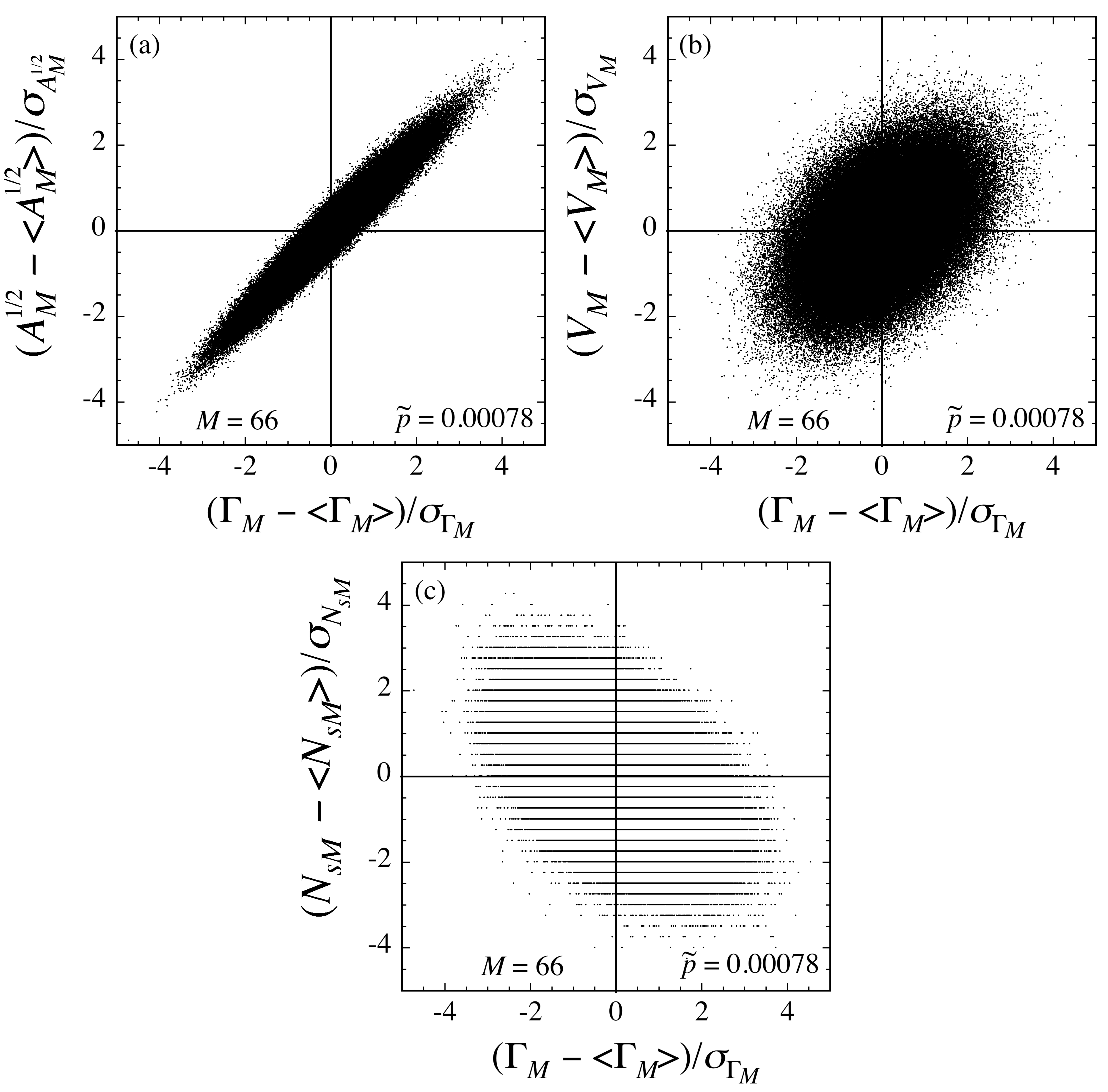}
\caption{Scatter plots showing configuration specific values of $(\Gamma_M-\langle\Gamma_M\rangle)/\sigma_{\Gamma_R}$ vs $(X_M-\langle X_M\rangle)/\sigma_{X_M}$ for $X_M$ equal to the (a) square root of the force-tile area $A_M^{1/2}$; (b) Voronoi volume $V_M$; (c) number of small particles $N_{sM}$.  Here $\sigma_{X_M}$ is the standard deviation of variable $X_M$ and results are for the specific case $M=66$ and $\tilde p=0.00078$ (a cluster with $M=66$ has an average radius of $\langle R\rangle\approx 5.4$).
}
\label{f16}
\end{center}
\end{figure}

\begin{figure}[h]
\begin{center}
\includegraphics[width=2.8in]{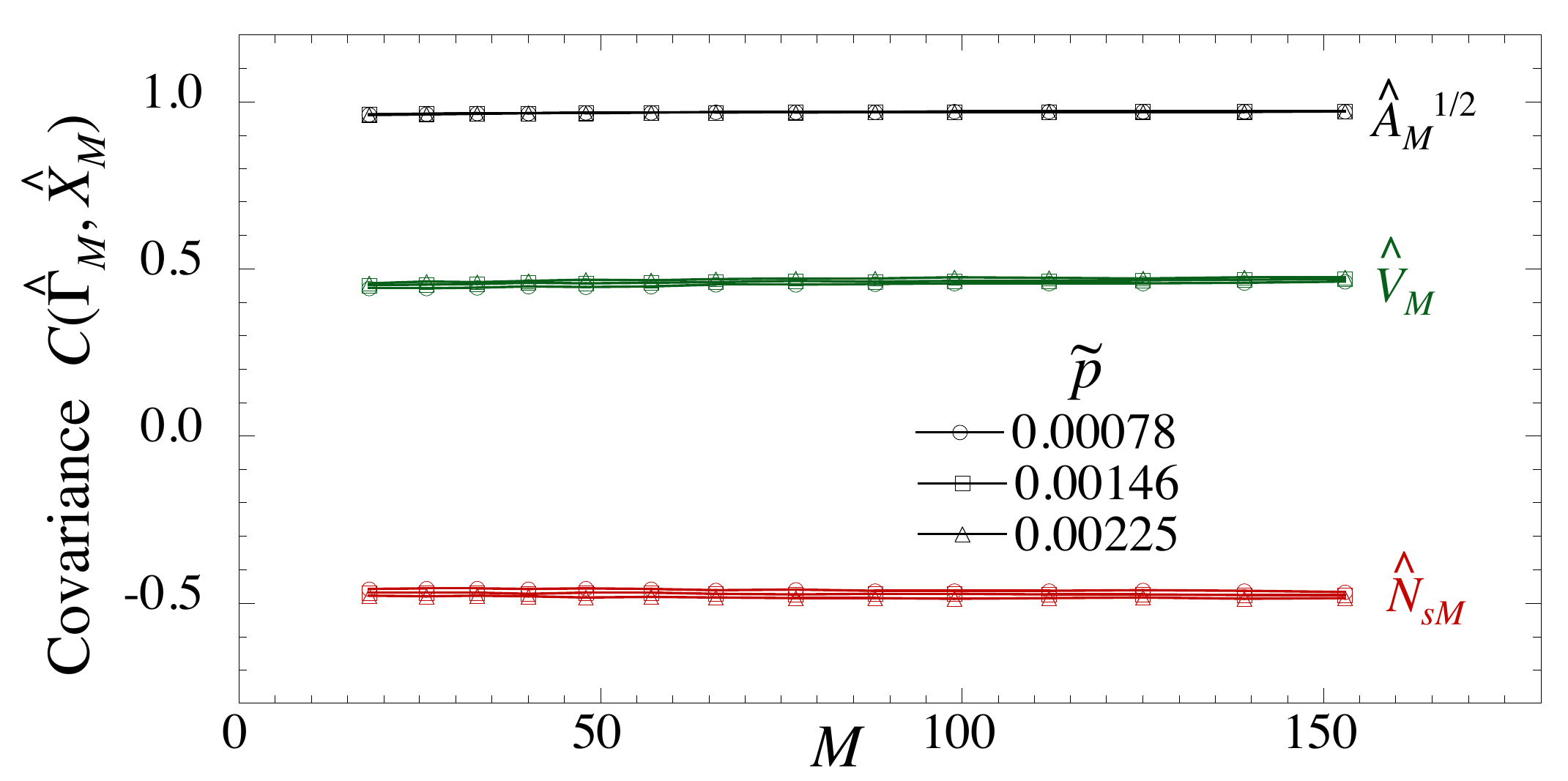}
\caption{(color online) Covariance between rescaled stress $\hat\Gamma_M$ and other variables vs the number of particles in the cluster $M$.  Results are shown for three different values of total system stress per particle $\tilde p$, as indicated by different symbol shapes.  The rescaled variables are defined by $\hat X_M\equiv (X_M-\langle X_M\rangle)/\sigma_{X_M}$, with $\sigma_{X_M}$ the standard deviation of $X_M$, and the plot shows results for $X_M$ equal to the square root of the force-tile area $A_M^{1/2}$, Voronoi volume $V_M$, and number of small particles $N_{sM}$. Our system has a total of $N=8192$ particles.
}
\label{f17}
\end{center}
\end{figure}

Finally we consider the correlations between the different conserved quantities in the clusters with fixed number of particles $M$.  We use the rescaled variables $\hat X_M$ defined similarly as in Eq.~(\ref{hatXR}).
First we consider the correlations with the stress $\hat\Gamma_M$.  In Fig.~\ref{f16} we show
scatter plots of the configuration specific values of $\hat \Gamma_M$ vs the other variables, for the particular case of $M=66$ (with average cluster radius $\langle R\rangle\approx 5.4$) and $\tilde p=0.00078$. 
In Fig.~\ref{f17} we plot the covariance between $\hat\Gamma_M$ and the other variables vs $M$, for three different values of the total system stress per particle $\tilde p$.  
As was observed for the clusters of fixed radius $R$, we
see that the covariances are essentially independent of $\tilde p$.  The correlation of $\hat\Gamma_R$ with $\hat A_R^{1/2}$ is again very close to  the maximum value of unity.  Correlations with $V_M$ and $N_{sM}$ are both roughly 50\%, but of opposite sign.
But the most striking result is that the correlations now stay essentially constant as $M$ increases, rather than decreasing with increasing cluster size as was observed for the clusters of constant radius $R$ in Fig.~\ref{f9}.

\begin{figure}[h]
\begin{center}
\includegraphics[width=2.8in]{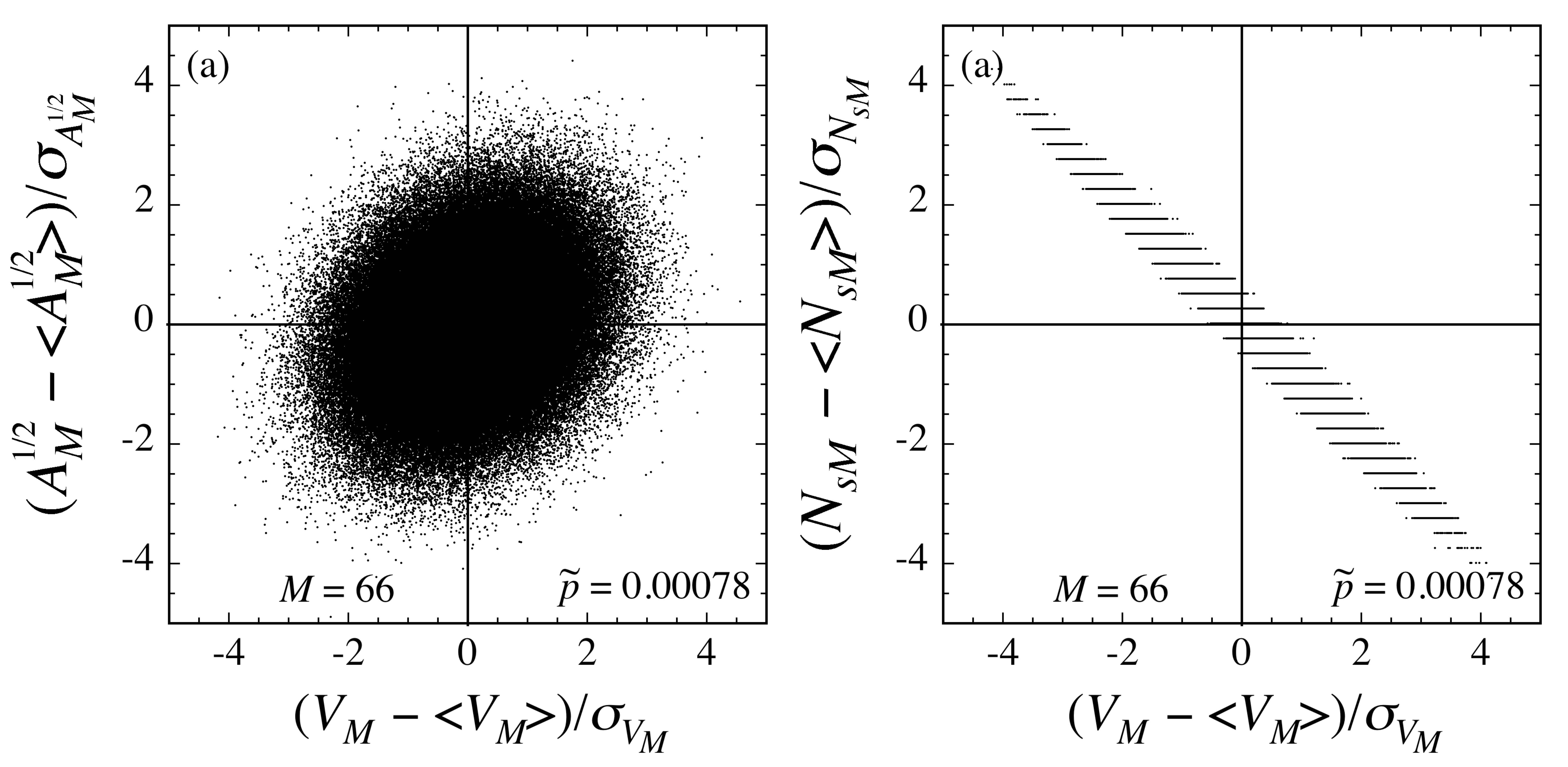}
\caption{Scatter plots showing configuration specific values of $(V_M-\langle V_M\rangle)/\sigma_{V_R}$ vs $(X_M-\langle X_M\rangle)/\sigma_{X_M}$ for $X_M$ equal to the (a) square root of the force-tile area $A_M^{1/2}$; (b) number of small particles $N_{sM}$.  Here $\sigma_{X_M}$ is the standard deviation of variable $X_M$ and results are for the specific case $M=66$ and $\tilde p=0.00078$ (a cluster with $M=66$ has an average radius of $\langle R\rangle\approx 5.4$).
}
\label{f18}
\end{center}
\end{figure}

\begin{figure}
\begin{center}
\includegraphics[width=2.8in]{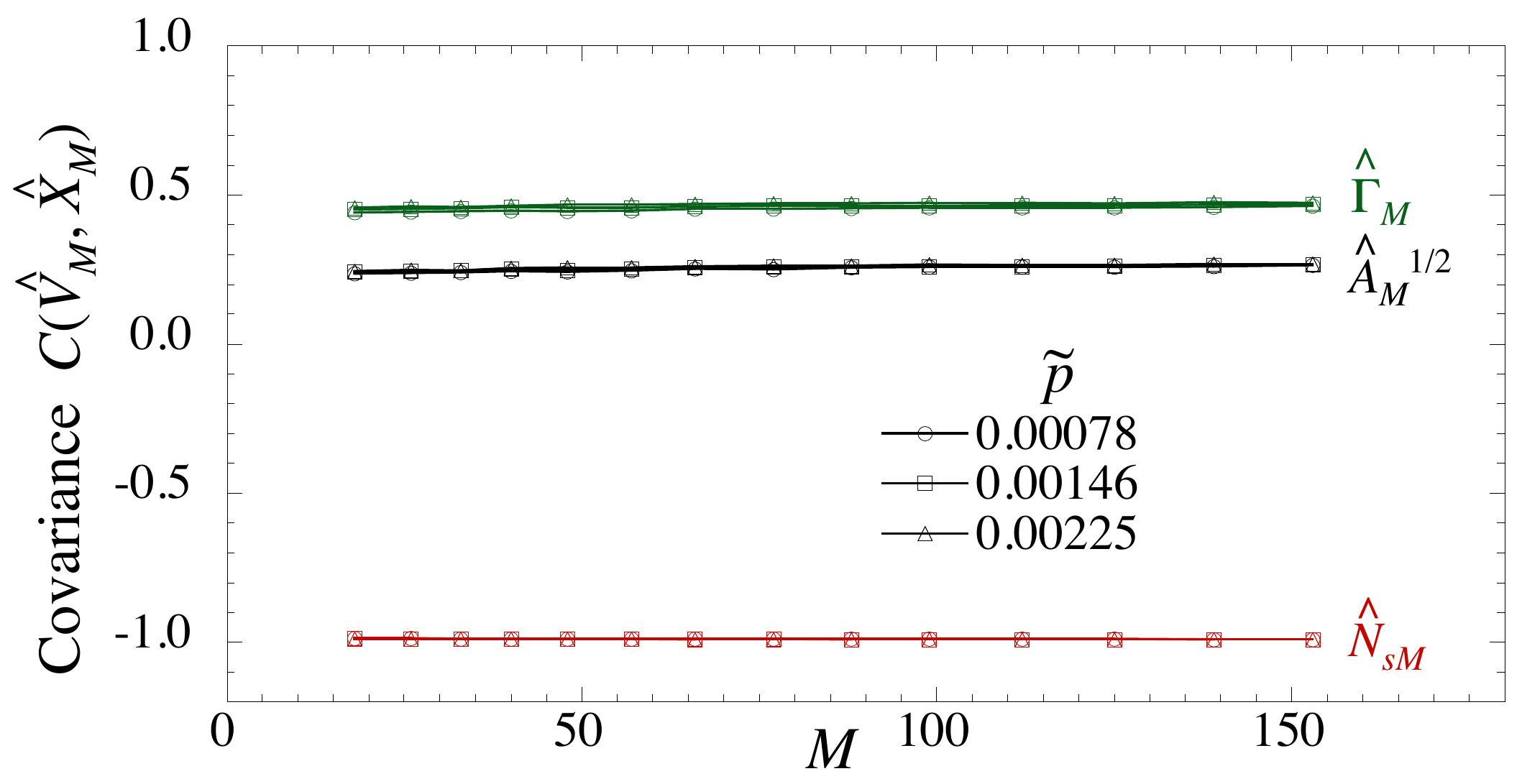}
\caption{(color online) Covariance between rescaled stress $\hat V_M$ and other variables vs the number of particles in the cluster $M$.  Results are shown for three different values of total system stress per particle $\tilde p$, as indicated by different symbol shapes.  The rescaled variables are defined by $\hat X_M\equiv (X_M-\langle X_M\rangle)/\sigma_{X_M}$, with $\sigma_{X_M}$ the standard deviation of $X_M$, and the plot shows results for $X_M$ equal to the stress $\Gamma_M$, square root of the force-tile area $A_M^{1/2}$ and number of small particles $N_{sM}$. Our system has a total of $N=8192$ particles.
}
\label{f19}
\end{center}
\end{figure}

We now consider the correlations with the Voronoi volume $\hat V_M$. In Fig.~\ref{f18} we show
scatter plots of the configuration specific values of $\hat V_M$ vs the other variables, for the particular case of $M=66$ and $\tilde p=0.00078$. 
In Fig.~\ref{f19} we plot the covariance between $\hat V_M$ and the other variables vs $M$, for three different values of the total system stress per particle $\tilde p$.  
The covariances are again essentially independent of $\tilde p$.  The correlation of $\hat V_R$ with $\hat N_{sM}$ is the strongest, close to the maximum magnitude of unity, but with negative sign (anti-correlated).  
Again, the correlations  stay essentially constant as $M$ increases.

\section{Discussion}

In this work we have considered mechanically stable packings of soft-core, frictionless, bidispersive disks in 2D, above the jamming transition.  Our packings are restricted to those having an isotropic total stress tensor.  We measure the statistical behavior of conserved quantities defined on clusters  ${\cal C}$ of fixed radius $R$, and clusters of
fixed number of particles $M$.  For conserved quantities we have considered the stress $\Gamma_{\cal C}$, defined as 1/2 the trace of the stress tensor, the Maxell-Cremona force-tile area $A_{\cal C}$, the Voronoi volume $V_{\cal C}$, the total number of particles $N_{\cal C}$ and the number of small particles $N_{s{\cal C}}$.  We have computed their averages, variances, and the correlations between them as a function of cluster size and the stress per particle of the total system $\tilde p = \Gamma_N/N$.

We find striking differences in the behavior of the two different ensembles of clusters.  For clusters with fixed radius $R$, average values of quantities defined on the cluster are simply determined from the corresponding known value for the entire system, $\langle X_R\rangle = X_N(\pi R^2/V)$, for all values of $R$.
In particular, the average Voronoi volume $\langle V_R\rangle$ is just the circle volume $\pi R^2$, and the relative fluctuations of $V_R$ are suppressed, scaling as $1/R^{3/2}$, in comparison the relative fluctuations of the other quantities, which scale as $1/R$.  Correlations are very strong between stress $\Gamma_R$ and force-tile area $A_R$, but correlations between $\Gamma_R$ and the other variables decay as the cluster size $R$ increases.  

For clusters with fixed number of particles $M$, however, the average $\langle X_M\rangle$ only algebraically approaches the naively expected value $X_N(M/N)$ as the cluster size $M$ increases.  The average on a finite cluster, therefore, is not apriori known without obtaining further information about the system beyond the values of its global parameters.  More strikingly, correlations between all pairs of conserved quantities appear to remain constant as the cluster size $M$ increases.

These results lead to our main conclusion, that for describing the stress distribution within such jammed packing, the cluster ensemble at fixed radius $R$ appears much more promising for use with maximum entropy models; one need only consider the two strongly correlated variables $\Gamma_R$ and $A_R$, as correlations with other variables will decrease as the cluster size increases.  Indeed, we have recently carried out just such an analysis \cite{WuTeitel} and have found good results.  For analyses based on clusters with a fixed number of particles $M$, it may be necessary to keep track of all conserved quantities, since correlations do not seem to decay with increasing cluster size, and these correlations are not in general small.

We have also made several other interesting observations: (i) 
We find that hyperuniformity, previously observed for packings exactly at the jamming $\phi_J$, continues to hold in packings above $\phi_J$. (ii) We find, in our bidisperse system, that the average concentration of small particles in a cluster is not uniform, but rather approaches the global value algebraically as the cluster size increases.  We find this for both clusters of fixed radius, and clusters of fixed number of particles.

\section*{Acknowledgments}

This work was supported by NSF Grant No. DMR-1205800.  Computations were carried out at the Center for Integrated Research Computing at the University of Rochester.  We wish to thank B.~Chakraborty, C.~E.~Maloney, B.~P.~Tighe and D.~V{\aa}gberg for helpful discussions.


\begin{thebibliography}{99}
\bibitem{note2}Given the rather limited range of $\tilde p$ that we study, the exact numerical values of the exponents reported here should not be taken as too definitive; they should not be viewed as critical exponents describing the true critical behavior as $\phi\to\phi_J$.
\end{thebibliography}
\end{document}